\documentclass[superscriptaddress, reprint, amsmath, amssymb, aps, pra, floatfix,footinbib]{revtex4-2}

\usepackage{bm}
\usepackage{algpseudocode}
\usepackage{algorithm}
\usepackage{booktabs}
\usepackage{mdframed}

\usepackage{mathtools}
\usepackage{amsfonts}
\usepackage{amssymb}
\usepackage{dsfont}
\usepackage{amsthm}

\usepackage[english]{babel}

\usepackage{CJKutf8}

\usepackage{graphicx}
\usepackage{amsmath}
\usepackage{lipsum}
\usepackage{wrapfig}
\usepackage{float}
\usepackage{enumitem} 
\usepackage{color}
\usepackage{hyperref}
\usepackage[normalem]{ulem}
\usepackage{url}
\usepackage{longtable}

\usepackage{comment}

\usepackage{array}
\usepackage{subfloat}

\useunder{\uline}{\ul}{}

\hypersetup{colorlinks = true}

\def\bra#1{\ensuremath{\mathinner{\langle{#1}|}}}
\def\ket#1{\ensuremath{\mathinner{|{#1}\rangle}}}

\newcommand{\braket}[2]{\langle #1|#2\rangle}

\newcommand{\needcite}[1]{\textcolor{red}{[Ref needed]}}
\usepackage{qcircuit}

\newcommand{\Leiden}{\affiliation{%
		$\langle aQa^{L}\rangle$ Applied Quantum Algorithms, Leiden University, Netherlands}\affiliation{%
		LIACS, Leiden University, Netherlands}}
\newcommand{\ASU}{\affiliation{School of Electrical, Computer, and Energy Engineering, Arizona State University, Tempe, Arizona 85281, USA}}
\begin{document}

\title{Equating quantum imaginary time evolution, Riemannian gradient flows, and stochastic implementations}

\author{Nathan A. McMahon}\email{n.a.mcmahon@liacs.leidenuniv.nl}\Leiden
\author{Mahum Pervez}\ASU
 \author{Christian Arenz}\email{carenz1@asu.edu}\ASU

\begin{abstract}
We identify quantum imaginary time evolution as a Riemannian gradient flow on the unitary group. 
We develop an upper bound for the error between the two evolutions that can be controlled through the step size of the Riemannian gradient descent which minimizes the energy of the system. We discuss implementations through adaptive quantum algorithms and present a stochastic Riemannian gradient descent algorithm in which each step is efficiently implementable on a quantum computer. We prove that for a sufficiently small step size, the stochastic evolution concentrates around the imaginary time evolution, thereby providing performance guarantees for cooling the system through stochastic Riemannian gradient descent.     
\end{abstract}

\maketitle

\emph{Introduction} -- Imaginary time evolution (ITE) is a powerful classical method for determining properties of a many-body Hamiltonian $H$. For example, various quantum Monte Carlo methods \cite{suzuki2012quantum} aim to 
find the spectral characteristics of $H$ by solving the time imaginary Schrödinger equation \cite{goldberg1967integration} whose normalized solution is given by the state 
\begin{align}
\label{eq:ITE_State}
\ket{\psi(\beta)}=\frac{e^{-\beta H}\ket{\psi_{0}}}{\Vert e^{-\beta H}\ket{\psi_{0}}\Vert },
\end{align}
where $\beta$ is the ``imaginary time''  and $\ket{\psi_{0}}$ is the initial state of the system. The appeal of such methods is the fact that if the initial state has non-zero overlap with the ground state $\ket{E_{0}}$ of $H$, i.e. $\bra{\psi_{0}}E_{0}\rangle\neq 0$, then under sufficiently long evolution times $\beta$ the state $\ket{\psi(\beta)}$ is given by \ket{E_{0}}.

Due to the exponential overhead in implementing imaginary time evolution on classical computers, over the last years there has been a growing interest in developing quantum algorithms that aim to create the ITE state \eqref{eq:ITE_State} on a quantum computer. Such methods face the challenge that the evolution $e^{-\beta H}$ that leads to the state \eqref{eq:ITE_State} is not unitary, and therefore not directly implementable with unitary gates on a quantum device. Several approaches \cite{motta2020determining, yeter2020practical, gomes2020efficient, sun2021quantum, lin2021real, nishi2021implementation,yeter2022quantum, tsuchimochi2023improved, liu2021probabilistic, kosugi2022imaginary, leamer2024quantum} have been proposed to overcome this challenge by designing unitary evolutions that approximate $\ket{\psi(\beta)}$. For instance, the quantum imaginary time evolution  algorithm \cite{motta2020determining, yeter2020practical, gomes2020efficient, sun2021quantum,nishi2021implementation, lin2021real, yeter2022quantum,tsuchimochi2023improved} and its probabilistic modifications \cite{liu2021probabilistic, kosugi2022imaginary}  Trotterize ITE to approximate each Trotter step by a unitary that is identified through classical optimization. Similarly, in variational approaches to ITE \cite{mcardle2019variational,jones2019variational} classical optimization routines are used in conjunction with a quantum computer to optimize a parameterized quantum circuit that prepares $\ket{\psi(\beta)}$. While these algorithms are often powerful heuristics, quantum algorithms that implement ITE with performance guarantees remain scarce in the literature.

\begin{figure}[t]
\centering
\includegraphics[width=0.85\columnwidth]{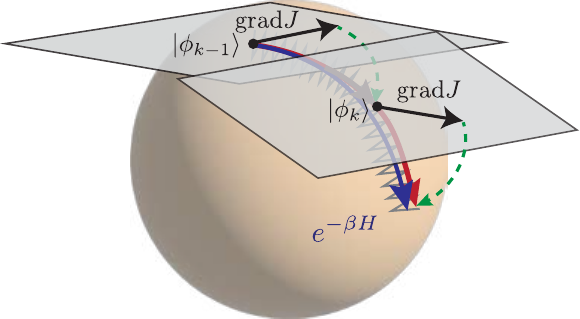}
 \caption{\label{fig:Intro} Schematic representation of the correspondence between imaginary time evolution (ITE) described by $e^{-\beta H}$ and Riemannian gradient descent (RGD) minimizing the energy of the system. A unitary evolution $e^{-\Delta \beta \text{grad}J}$ (red) is created by the retraction (green) of the Riemannian gradient $\text{grad}J$ with respect to a cost function $J$ where $\Delta\beta$ is the step size. We show that the error between the ITE state and the state $\ket{\phi_{1}}$ created through one step of RGD is of the order $\mathcal O(\Delta \beta^{-2})$ (Lemma 1). Building upon this, we prove that after $n$ steps the state $\ket{\phi_{n}}$ approximates the ITE state with an error $\mathcal O(\Delta\beta)$ (Theorem 1). We use this observation to develop a stochastic Riemannian gradient descent algorithm (gray line) whose evolution concentrates around ITE (Theorem 2). }
\end{figure}

In this work, we address this challenge by identifying ITE as a Riemannian gradient flow \cite{helmke2012optimization,schulte2010gradient,malvetti2024randomized, wiersema2023optimizing,   magann2023randomized} on the unitary group, which can be implemented through adaptive quantum algorithms \cite{grimsley2019adaptive, tang2021qubit, magann2022feedback, larsen2024feedback,magann2023randomized, magann2022lyapunov, wiersema2023optimizing, tang2024non} on quantum computers. The relationship between ITE and gradient flows has been observed before in the literature \cite{stokes2023numerical} and has recently been leveraged \cite{gluza2024double} to provide fidelity bounds for ground state preparation and energy minimization. Here, we derive rigorous bounds for implementing the \emph{full} ITE, thus providing performance guarantees for implementing the ITE state for generic imaginary evolution times $\beta$ and for convergence to excited states (when $\braket{\psi_{0}}{E_{0}} = 0$). 
 
We first prove that the state created by the unitary evolution that describes the Riemannian gradient flow can be made arbitrarily close to the ITE state $\ket{\psi(\beta)}$. This is achieved by appropriately choosing the step size $\Delta\beta$ of the discretized Riemannian gradient flow evolution, which can be interpreted as a Riemannian gradient descent algorithm minimizing the energy of the system, as depicted in Fig. \ref{fig:Intro}. We go on to develop an upper bound (Theorem 1) for the error between the ITE state and the state that is created through Riemannian gradient descent, which can again be controlled through $\Delta \beta$. The developed equivalence between ITE and Riemannian gradient flows has two key implications: (i) it motivates using the large literature on heuristic adaptive quantum algorithms \cite{grimsley2019adaptive, tang2021qubit, magann2022feedback, larsen2024feedback,magann2023randomized, magann2022lyapunov, wiersema2023optimizing, tang2024non} to design quantum algorithms for implementing ITE and (ii) the developed upper bound gives performance guarantees for state preparation methods that are based on Riemannian gradient flows \cite{wiersema2023optimizing, magann2023randomized, malvetti2024randomized,gluza2024double_2}. 

We demonstrate the utility of (i) and (ii) by developing a stochastic implementation of the discretized Riemannian gradient flow. Each step in this stochastic Riemannian gradient descent algorithm is efficiently implementable on a quantum computer and the energy of the system is provably reduced on average in each step. We then derive performance bounds for using the stochastic Riemannian gradient descent algorithm to prepare the ITE state $\ket{\psi(\beta)}$ to arbitrary precision (Theorem 2).  We show that the randomized evolution concentrates around ITE, converging at a rate determined by $\Delta \beta$. We proceed to use the developed bounds to obtain insights on how knowledge of the initial state and $H$ may be exploited to develop quantum circuit implementations that prepare the ITE state efficiently.

\emph{Quantum imaginary time evolution as Riemannian gradient descent} -- The theory of Riemannian gradient flows is a rich framework for solving optimization problems defined on a Riemannian manifold, such as the unitary group \cite{helmke2012optimization}. Riemannian gradient flows have a long history in optimization. Applications range from diagonalizing a matrix through Brocket's double bracket flow to solving least squares type problems \cite{brockett1989least,brockett1993differential} and finding the ground state of a many-body Hamiltonian \cite{dawson2008unifying, hastings2022lieb}. Recently, Riemannian gradient flows have been utilized to develop quantum algorithms for ground state problems \cite{wiersema2023optimizing,magann2023randomized,malvetti2024randomized}. Rather than optimizing over parameters in a quantum circuit to minimize cost function $J$ (e.g., as in variational quantum algorithms \cite{cerezo2021variational}), $J$ is instead directly optimized over unitary transformations.

The Riemannian gradient flow on the special unitary group $\text{SU}(d)$ of dimension $d$ is defined by the solution to the differential equation \cite{helmke2012optimization,schulte2010gradient} 
\begin{align}
\label{eq:gradflowdef}
\frac{d}{dt}U=-\text{grad} J[U]
\end{align}
for the unitary operator $U\in \text{SU}(d)$. Here, $\text{grad} J[U] \in T_{U}\text{SU}(d) $ is the Riemannian gradient  of a cost function $J[U]$ that lives in the tangent space $T_{U}\text{SU(d)}$ given at the identity $T_{\mathds{1}}\text{SU(d)}=\mathfrak{su}(d)$ by the special unitary algebra $\mathfrak{su}(d)$. In this work we consider a cost function that is given by the expectation value $J=\bra{\phi} H\ket{\phi}$ of the Hamiltonian $H$ with respect to the state $\ket{\phi}=U\ket{\psi_{0}}$. This state is created by applying the unitary transformation $U$ to the initial state $\ket{\phi_{0}}$. The choice of the cost function is motivated by the fact that the solution to \eqref{eq:gradflowdef} prepares the groundstate of $H$,  by minimizing $J$, when $t$ is sufficiently large \cite{helmke2012optimization,schulte2010gradient}. 

The discretized solution to the differential equation \eqref{eq:gradflowdef} defining the Riemannian gradient flow is given by a sequence of unitary transformations of the form 
\begin{align}
\label{eq:updatestep}
U(\Delta t)=e^{-\Delta t \text{grad J}},
\end{align}
where $\Delta t$ is the step size and $\text{grad}J=[H,\ket{\phi}\bra{\phi}]\in\mathfrak{su}(d)$ is the Riemannian gradient at the state $\ket{\phi}$ \footnote{With a slight abuse of notation, from now on we denote the Riemannian gradient by $\text{grad}J= [H,\ket{\phi}\bra{\phi}]$, noting that due to the invariance of the Hilber-Schmidt inner product $\langle X,Y\rangle=\text{Tr}\{X^{\dagger}Y\}$ with respect to $U$, the inner product between tangent space elements $X,Y\in T_{U}\text{SU}(d)$ and the inner product between elements $X,Y\in\mathfrak{su}(d)$ is the same}.

The state that is created by successively applying the transformation \eqref{eq:updatestep} $k$-times to the initial state $\ket{\phi_{0}}=\ket{\psi_{0}}$ is recursively given by
\begin{align}
\label{eq:discretizedRec}
\ket{\phi_{k}}=U_{k}(\Delta t)\ket{\phi_{k-1}}.
\end{align}
 The recursive update \eqref{eq:discretizedRec} can be understood as a \emph{Riemannian gradient descent} (RGD) algorithm that minimizes $J$.  We see that the unitary transformation $U_{k}(\Delta t)=e^{-\Delta t\text{grad}J}$ where $\text{grad}J=[H,\ket{\phi_{k-1}}\bra{\phi_{k-1}}]$ creating the state $\ket{\phi_{k}}$ depends on the previous state  $\ket{\phi_{k-1}}$. This observation indicates that designing a quantum algorithm that implements RGD requires some ``feedback'' mechanism that leverages the information about the previous state to inform the update step. 

Before we discuss how a quantum algorithm can be designed that adaptively grows a quantum circuit to create $\ket{\phi_{n}}$ based on measurement data, we first discuss how RGD can be used to create the ITE state \eqref{eq:ITE_State} arbitrarily well. 
We observe that for small step sizes $\Delta t=\Delta \beta$, the ITE state $\ket{\psi(\Delta\beta)}$ and the state $\ket{\phi(\Delta t)}=U_{1}(\Delta t)\ket{\psi_{0}}$ created through one step of RGD are close to each other, which is expressed through the relation
$\left.\frac{d}{dx}\ket{\psi(x)}\right|_{x=0}=\left.\frac{d}{dx}\ket{\phi(x)}\right|_{x=0}$.
As such, the Euclidean norm difference $\Vert \ket{\psi(\Delta \beta)}-\ket{\phi(\Delta \beta)} \Vert$  between the two states is of the order $\mathcal O(\Delta \beta^{2})$. We remark that this observation is equivalent to the observation made in \cite{gluza2024double} that the ITE state solves Brocket's double bracket flow equation \cite{brockett1989least, helmke2012optimization}, which describes the Riemannian gradient flow on the adjoint orbit of the unitary group.  

An upper bound can be obtained for $\Vert \ket{\psi(\Delta \beta)}-\ket{\phi(\Delta \beta)} \Vert$ by bounding the remainder of the corresponding Tailor expansions. 
With further details found in Appendix \ref{sec:Lemma1Theorem1}, we establish the following Lemma.

\textbf{Lemma 1}: \textit{Let}
$\ket{\psi(\Delta\beta)}$ and $ \ket{\phi(\Delta \beta)}=U(\Delta \beta)\ket{\psi_{0}}$ 
\textit{be the states created by ITE \eqref{eq:ITE_State} and one step of RGD \eqref{eq:updatestep} with step size $\Delta \beta$.  
 Then for any initial state $\ket{\psi_{0}}$, }
 \begin{align}
 \Vert \ket{\psi(\Delta \beta)}-\ket{\phi(\Delta \beta)} \Vert \leq  6\Delta\beta^{2}\Vert H\Vert_{\infty}^{2},
 \end{align}
\textit{where $\Vert H\Vert_{\infty}$ denotes the spectral norm of $H$}.

Consequently, if we divide $\beta$ into $n$ segments $\Delta \beta=\frac{\beta}{n}$, we expect that the error $\epsilon_{n}=\Vert\ket{\psi(\beta)}-\ket{\phi_{n}}\Vert$ between the state $\ket{\phi_{n}}$ created through $n$ steps of RGD \eqref{eq:discretizedRec} and the ITE state  $\ket{\psi(\beta)}$ to be of the order $\mathcal O(1/n)$. 

The main technical challenge in rigorously establishing this intuition is the fact that standard techniques used to show that the total error $\epsilon_{n}$ is upper bounded by $n$ times the local error (i.e., bounded by Lemma 1) cannot be directly applied, due to the non-unitary nature of ITE.  Instead, we show in the Appendix \ref{sec:Lemma1Theorem1} that $\epsilon_{k}$ is given by a recursive relation of the form 
\begin{align}
\label{eq:recursionRel}
\epsilon_{k}\leq \epsilon_{k-1}A+B
\end{align}
where $A=1+4\Delta \beta \Vert H\Vert_{\infty}$ and $B=10\Delta \beta^{2}\Vert H\Vert_{\infty}^{2}$. As such, the error $\epsilon_{n}$ is upper bounded by a geometric series
\begin{align}
\epsilon_{n}\leq B\sum_{k=0}^{n-1}A^{k}= B \frac{1-A^{n}}{1-A},  
\end{align}
which establishes the following bound.

\textbf{Theorem 1}: \textit{The error $\epsilon_{n}$ between the ITE state $\ket{\psi(\beta)}$ \eqref{eq:ITE_State} and the state $\ket{\phi_{n}}$ \eqref{eq:discretizedRec} created by $n$ steps of RGD with step size $\Delta \beta=\frac{\beta}{n}$ is upper bound by}
\begin{align}
\label{eq:boundTH}
\epsilon_{n}\leq \frac{5}{2} \Delta \beta \Vert H\Vert_{\infty}\left(e^{4\beta \Vert H\Vert_{\infty}}-1\right). 
\end{align}
Thus, for $\beta \Vert H  \Vert_{\infty}\ll 1$, 
such that the second order of $e^{\beta \Vert H\Vert_{\infty}}$ can be neglected, the complexity for preparing the corresponding ITE state through RGD is of the order $\mathcal O(\frac{\beta^{2}\Vert H\Vert_{\infty}^{2}}{\epsilon})$. Such a short ITE has been used in primitive subroutines in other quantum algorithms, such as the quantum Lanczos or the quantum minimally entangled typical thermal states algorithm \cite{white2009minimally,motta2020determining}. Thus, Theorem 1 provides error guarantees for the implementation of these subalgorithms via RGD.

We further remark that since by the triangle inequality we have $\Vert \ket{E_{0}}-\ket{\phi_{n}} \Vert\leq \Vert \ket{E_{0}}-\ket{\psi(\beta)} \Vert +\epsilon_{n}$, Theorem 1 allows us to bound the error for preparing the ground state $\ket{E_{0}}$ of $H$ via RGD \cite{wiersema2023optimizing, magann2023randomized, malvetti2024randomized}. However, Theorem 1 goes beyond ground state preparation as it bounds the error for preparing a generic ITE state. This includes preparing excited states, which can be achieved by initializing the system in a state that has zero overlap with the ground state.

\emph{Gradient flow-based quantum algorithms for quantum imaginary time evolution} -- The RGD update step $U_{k}=e^{-\Delta \beta \text{grad}J}$ is, in general, not efficiently implementable on a quantum computer. As the Riemannian gradient $\text{grad}J\in\mathfrak{su}(2^{N})$ generally lives in an exponentially large Lie algebra $\mathfrak{su}(2^{N})$ of dimension $2^{2N}-1$ where $N$ is the number of qubits, we can in general not hope for implementing $U_{k}$ with polynomially many gates. However, several approximation schemes have been introduced in the literature that project the Riemannian gradient into smaller dimensional subspaces to make the update step efficiently implementable, e.g., through Trotterization \cite{grimsley2019adaptive, wiersema2023optimizing, magann2023randomized, malvetti2024randomized}. The key idea is to pick a polynomially sized subspace $\mathcal A_{k}\subset\mathfrak{su}(2^{N})$ in each update step $k$. The Riemannian gradient is then projected into this subspace to obtain an approximate Riemannian gradient 
$\widetilde{\text{grad}J}=i\sum_{P_{j}\in\mathcal A_{k}} C_{k,j} P_{j}$ 
in which the coefficients $C_{k,j}$ can be estimated via measurements of the gradients 
\begin{align}
\label{eq:coefficients}
C_{k,j}&=\left.\frac{\partial }{\partial \theta}\bra{\phi_{k}}e^{i\theta P_{j}}He^{-i\theta P_{j}}\ket{\phi_{k}}\right|_{\theta=0} \\ \nonumber 
&=\langle \text{grad}J,iP_{j}\rangle, 
\end{align}
e.g., through the parameter shift rule or finite differences \cite{mitarai2018quantum, schuld2019evaluating}. Here, $\langle X,Y\rangle=\text{Tr}\{X^{\dagger}Y\}$ denotes the Hilbert-Schmidt inner product.   

A particularly appealing strategy is to project into a randomly chosen directions $P_{j}$ (in the tangent space) in each update step \cite{magann2023randomized}. For ground state problems, this randomization strategy convergences to the ground state almost surely despite the existence of saddle points \cite{malvetti2024randomized}. Here, we employ a similar randomization strategy to efficiently implement $U_{k}$ \emph{on average}.

Consider sampling in each update step $k$ a gate $V_{k}^{(j)}=e^{-i C_{k,j}D\Delta \beta P_{j}}$ by picking uniformly random a (normalized) Pauli operator $iP_{j}\in\mathfrak{su}(d)$ where $D=\text{dim}(\mathfrak{su}(d))=d^{2}-1$. We recall that the coefficients $C_{k,j}$ are determined by \eqref{eq:coefficients} and depend on the state of the previous step. To gain some intuition on how this randomization procedure gives rise to implementing ITE, we start by considering the energy change $\Delta E_{j}=\bra{\phi}H\ket{\phi}-\bra{\phi}V_{k}^{(j)\dagger}HV_{k}^{(j)}\ket{\phi}$ at the step $k$ for some state $\ket{\phi}$. For a step size $\Delta \beta=1/(4D\Vert H\Vert_{\infty}) $ this change is lower bounded by 
$\Delta E_{j} \geq  \frac{1}{8\Vert H \Vert_{\infty} } \langle \text{grad}J, iP_{j}\rangle^{2}
$ \cite{magann2023randomized}. Sampling Pauli operators $P_{j}$ from $\mathfrak{su}(d)$ uniformly gives, for any $A$, that $\frac{1}{D}\sum_{j}P_{j}AP_{j}=\frac{\text{Tr}\{A\}}{d}$. As such, we find that the average energy change $\overline{\Delta E}=\mathbb E_{j}[\Delta E_{j}]$ at a single random step $k$ is lower bound by 
\begin{align}
\label{eq:averageenergychange}
\overline{\Delta E} \geq \frac{\Vert \text{grad}J\Vert_{\text{HS}}^{2}}{8 d\Vert H\Vert_{\infty}}.
\end{align}
The Hilbert-Schmidt norm $\Vert \cdot \Vert _{\text{HS}}$ of the Riemannian gradient $\text{grad}J=[H,\ket{\phi}\bra{\phi}]$ is given by the variance  
$\Vert\text{grad}J\Vert_{\text{HS}}=\sqrt{\bra{\phi}H^{2}\ket{\phi}- \bra{\phi}H\ket{\phi}^{2}}$, which is zero only at eigenstates of the Hamiltonian. As such, on average a randomized step corresponds to the energy change that is obtained for one step of RGD. In fact, to first order in the step size, the averaged evolution is described by the unitary quantum channel that corresponds to one step of RGD.

\begin{figure}
\centering
\includegraphics[width=0.9\columnwidth]{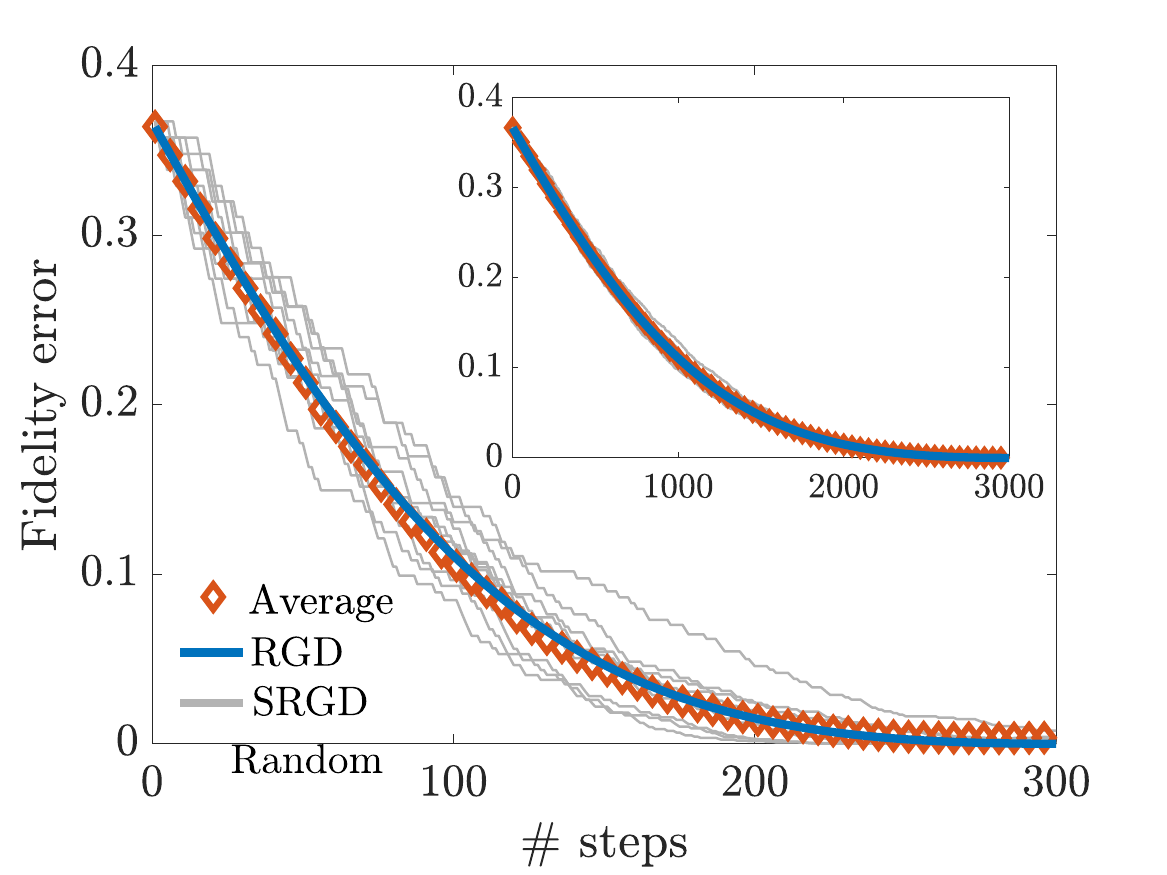}
 \caption{\label{fig:Fig1} Fidelity error between the ITE state \eqref{eq:ITE_State} with $\beta=1$ and the state \eqref{eq:discretizedRec} created by Riemannian gradient descent (blue) for a single qubit with initial state $\ket{\psi_{0}}=\ket{+}$ and Hamiltonian $H=Z$ given by the Pauli-Z operator. Plotted as a function of the number of steps $k$ for a step size $\Delta\beta=\beta/n$ where $n=300$. The grey lines show the error obtained from stochastic Riemannian gradient descent implemented by projecting the Riemannian gradient in each step into a uniformly random tangent space direction. The red diamonds show the average over $50$ trajectories (only $10$ are shown). In the inset plot the fidelity error is shown for $n=3000$.} 
\end{figure}

In Fig. \ref{fig:Fig1} we investigate how close a single random trajectory (grey curves) is to the ITE state, characterized by the fidelity error
\begin{align}
\label{eq:Fiderror}
\varepsilon_{k}^{(\gamma)}=1-| \bra{\psi(\beta)} \chi^{(\gamma)}_{k}\rangle|^{2}.
\end{align}
Here, $\ket{\chi^{(\gamma)}_{k}}$ is a particular random state generated by a sequence of random unitaries, $V_{k}^{(j)}$, corresponding to the choice of random Pauli operators $\gamma = \{iP_{1}, iP_{2}, \cdots, iP_{k}\}$. We refer to the sequence of random unitaries $V_{k}^{(j)}$ that creates the states $\ket{\chi^{(\gamma)}_{k}}$ as \emph{stochastic Riemannian gradient descent} (SRGD) \cite{malvetti2024randomized,gutman2023coordinate}.

We observe that the average over $50$ random trajectories $\gamma$ (red diamonds, Fig. \ref{fig:Fig1}) follows the evolution obtained by RGD remarkably well. From the inset plot that shows the fidelity error for a smaller step size $\Delta\beta$, we also see that the variance of the fidelity error can be controlled through $\Delta \beta$. Indeed, the plot suggests that the smaller the step size $\Delta \beta$, the closer a random trajectory comes to the evolution obtained from RGD. This observation can be made precise through the following Theorem.

\textbf{Theorem 2:} \textit{The average fidelity error  $\bar{\varepsilon}_{n}=\mathbb E_{\gamma}[\varepsilon_{n}^{(\gamma)}]$ after $n$ steps of SRGD with step size $\Delta\beta=\frac{\beta}{n}$ is upper bounded by }
\begin{align}
    b_{n}=\frac{9}{2}\sqrt{2\Delta\beta\Vert H\Vert_{\infty}}D \left(e^{8\beta\Vert H\Vert_{\infty}}-1\right)^{\frac{1}{2}},
\end{align}
\textit{for sufficiently large $n$. For any $\delta>0$, the probability that a state created through SRGD will give rise to a fidelity error greater than $b_{n}+\delta$ is upper bounded by }
\begin{align}
    \text{Pr}(\varepsilon_{n}^{(\gamma)} > b_{n} + \delta)\leq \frac{8\Delta\beta \Vert H\Vert_{\infty}D^{2}}{\delta^{2}} \left(e^{8\beta \Vert H\Vert_{\infty}}-1\right)
    \end{align}

We prove Theorem 2 in Appendix \ref{app:Theorem2} by establishing a recursion relation for the variance similar to \eqref{eq:recursionRel} and then applying the Chebyshev inequality. 

Theorem 2 shows that by choosing the step size $\Delta\beta$, we can control the deviation $\delta$ from the average fidelity error upper bound $b_{n}$. Since $b_{n}$ can also be controlled in the same way, we see that a random state $\ket{\chi_{n}^{(\gamma)}}$ created through SRGD may be made arbitrarily close to the ITE state. This observation provides an interesting perspective of SRGD as ``cooling'' the initial state $\ket{\psi_{0}}$ to the ground state of $H$.

\emph{Discussion} -- Each step of SRGD is efficiently implementable on a quantum computer. The gates $V_{k}^{(j)}$, generated by Pauli operators, can be implemented by a quantum circuit of depth at most linear in the number of qubits. However, the overall runtime $\mathcal O(\beta \Vert H\Vert_{\infty}D^{2}\exp(8\beta \Vert H\Vert_{\infty}))$ of the randomized algorithm cannot be efficient in general for complexity reasons \cite{kempe2006complexity}, which is reflected in Theorem 2 by the dependence on $D=2^{2N}-1$.

In contrast, the lower bound for the average energy change \eqref{eq:averageenergychange} suggests that the complexity of SRGD scales inversely with the smallest coefficient $C_{j}=\langle \text{grad}J,iP_{j}\rangle^{2}$ of the Riemannian gradient (taken over all steps). Consequently, we expect that if there exists a constant $\alpha>0$ such that $C_{j}$ does not vanish faster than $\mathcal O(N^{-\alpha})$, the ITE state can be prepared in polynomially many steps. As the Riemannian gradient depends on the Hamiltonian $H$ and the state $\ket{\chi_{k}^{(\gamma)}}$, a more efficient sampling strategy should exploit additional properties of $H$ and $\ket{\chi_{k}^{(\gamma)}}$.

For example, consider the Riemannian gradient  $[H,\ket{\chi_{k}^{(\gamma)}}\bra{\chi_{k}^{(\gamma)}}]$ at a random state $\ket{\chi_{k}^{(\gamma)}}$ that has support on only a polynomially (in $N$) sized subset of Pauli operators at each step $k$. We then may instead sample only over this subset, giving rise to an effective sample space of dimension $D = \mathrm{poly}(N)$. In practice, we can only hope that the commutator  $[H,\ket{\chi_{k}^{(\gamma)}}\bra{\chi_{k}^{(\gamma)}}]$ is approximately supported on a polynomially sized subset of Pauli operators. We observe that such a scenario is possible if the state $\ket{\chi_{k}^{(\gamma)}}$ has finite correlation length, and $H$ is a geometrically local Hamiltonian. The finite correlation length suggests that expectation values of Pauli operators $P$, $\langle P\rangle = \bra{\chi_{k}^{(\gamma)}}P\ket{\chi_{k}^{(\gamma)}}$, will decay exponentially with the support of $P$. This implies that $\ket{\chi_{k}^{(\gamma)}}\bra{\chi_{k}^{(\gamma)}} \approx \sum_{P \in \mathcal{S}} \langle P \rangle P$ where $\mathcal{S}$ is a $\mathrm{poly}(N)$ sized set of all Pauli operators with support under some fixed constant.
In turn, this would imply that $[H,\ket{\chi_{k}^{(\gamma)}}\bra{\chi_{k}^{(\gamma)}}]$ would be well approximated using only $\mathrm{poly}(N)$ Pauli operators.

We also expect that the speed of convergence of RGD to the ITE state can be improved by utilizing higher order methods in which the step size $\Delta \beta$ becomes step dependent. 
For example, the step size $\Delta \beta$ may be chosen by employing second derivative information \cite{schulte2010gradient,kong2024quantitative}. In general, efficient approximations of the Riemannian gradient could also be informed by tensor network \cite{vanderstraeten2019tangent,miao2023quantum}, Bayesian, and machine learning approaches \cite{dunjko2018machine} that aim to learn the Riemannian gradient.

\emph{Conclusion} -- In this work we showed that quantum imaginary time evolution can be understood as Riemannian gradient descent that minimizes the energy of the system. We derived a rigorous bound for the error between the states created through imaginary time evolution and Riemannian gradient descent that can be controlled by the step size. We went on to prove bounds for stochastic implementations, in which each step of the Riemannian gradient descent algorithm is efficiently implementable on quantum computers. We characterized how much the energy decreases on average in a randomized step. Interestingly, due to a concentration result, the stochastic evolutions follows with high probability the imaginary time evolution. This implies that, for sufficiently small step sizes, the system is cooled by moving into random tangent space directions of the Riemannian gradient descent.

We discussed potential ways forward to leverage information about the Hamiltonian and the initial state to efficiently prepare the imaginary time state on a quantum computer. Problem dependent knowledge could inform when learning an efficient representation of the Riemannian gradient (i.e., in a lower dimensional subspace) may be possible. Ultimately, the derived bounds and discussions provide a jumping off point for utilizing the vast literature around adaptive quantum algorithms for performing imaginary time evolution on quantum computers.  \\

\emph{Acknowledgments} -- N.A.M. was co-funded by the European Union (ERC CoG, BeMAIQuantum, 101124342). M. P. and C.A. acknowledge support from the National Science Foundation (Grant No. 2231328).

\bibliography{References}

\begin{thebibliography}{44}%
\makeatletter
\providecommand \@ifxundefined [1]{%
 \@ifx{#1\undefined}
}%
\providecommand \@ifnum [1]{%
 \ifnum #1\expandafter \@firstoftwo
 \else \expandafter \@secondoftwo
 \fi
}%
\providecommand \@ifx [1]{%
 \ifx #1\expandafter \@firstoftwo
 \else \expandafter \@secondoftwo
 \fi
}%
\providecommand \natexlab [1]{#1}%
\providecommand \enquote  [1]{``#1''}%
\providecommand \bibnamefont  [1]{#1}%
\providecommand \bibfnamefont [1]{#1}%
\providecommand \citenamefont [1]{#1}%
\providecommand \href@noop [0]{\@secondoftwo}%
\providecommand \href [0]{\begingroup \@sanitize@url \@href}%
\providecommand \@href[1]{\@@startlink{#1}\@@href}%
\providecommand \@@href[1]{\endgroup#1\@@endlink}%
\providecommand \@sanitize@url [0]{\catcode `\\12\catcode `\$12\catcode
  `\&12\catcode `\#12\catcode `\^12\catcode `\_12\catcode `\%12\relax}%
\providecommand \@@startlink[1]{}%
\providecommand \@@endlink[0]{}%
\providecommand \url  [0]{\begingroup\@sanitize@url \@url }%
\providecommand \@url [1]{\endgroup\@href {#1}{\urlprefix }}%
\providecommand \urlprefix  [0]{URL }%
\providecommand \Eprint [0]{\href }%
\providecommand \doibase [0]{https://doi.org/}%
\providecommand \selectlanguage [0]{\@gobble}%
\providecommand \bibinfo  [0]{\@secondoftwo}%
\providecommand \bibfield  [0]{\@secondoftwo}%
\providecommand \translation [1]{[#1]}%
\providecommand \BibitemOpen [0]{}%
\providecommand \bibitemStop [0]{}%
\providecommand \bibitemNoStop [0]{.\EOS\space}%
\providecommand \EOS [0]{\spacefactor3000\relax}%
\providecommand \BibitemShut  [1]{\csname bibitem#1\endcsname}%
\let\auto@bib@innerbib\@empty
%</preamble>
\bibitem [{\citenamefont {Suzuki}(2012)}]{suzuki2012quantum}%
  \BibitemOpen
  \bibfield  {author} {\bibinfo {author} {\bibfnamefont {M.}~\bibnamefont
  {Suzuki}},\ }\href@noop {} {\emph {\bibinfo {title} {Quantum Monte Carlo
  Methods in Equilibrium and Nonequilibrium Systems: Proceedings of the Ninth
  Taniguchi International Symposium, Susono, Japan, November 14--18, 1986}}},\
  Vol.~\bibinfo {volume} {74}\ (\bibinfo  {publisher} {Springer Science \&
  Business Media},\ \bibinfo {year} {2012})\BibitemShut {NoStop}%
\bibitem [{\citenamefont {Goldberg}\ and\ \citenamefont
  {Schwartz}(1967)}]{goldberg1967integration}%
  \BibitemOpen
  \bibfield  {author} {\bibinfo {author} {\bibfnamefont {A.}~\bibnamefont
  {Goldberg}}\ and\ \bibinfo {author} {\bibfnamefont {J.~L.}\ \bibnamefont
  {Schwartz}},\ }\bibfield  {title} {\bibinfo {title} {Integration of the
  schr{\"o}dinger equation in imaginary time},\ }\href
  {https://doi.org/https://doi.org/10.1016/0021-9991(67)90049-6} {\bibfield
  {journal} {\bibinfo  {journal} {Journal of Computational Physics}\ }\textbf
  {\bibinfo {volume} {1}},\ \bibinfo {pages} {433} (\bibinfo {year}
  {1967})}\BibitemShut {NoStop}%
\bibitem [{\citenamefont {Motta}\ \emph {et~al.}(2020)\citenamefont {Motta},
  \citenamefont {Sun}, \citenamefont {Tan}, \citenamefont {O’Rourke},
  \citenamefont {Ye}, \citenamefont {Minnich}, \citenamefont {Brandao},\ and\
  \citenamefont {Chan}}]{motta2020determining}%
  \BibitemOpen
  \bibfield  {author} {\bibinfo {author} {\bibfnamefont {M.}~\bibnamefont
  {Motta}}, \bibinfo {author} {\bibfnamefont {C.}~\bibnamefont {Sun}}, \bibinfo
  {author} {\bibfnamefont {A.~T.}\ \bibnamefont {Tan}}, \bibinfo {author}
  {\bibfnamefont {M.~J.}\ \bibnamefont {O’Rourke}}, \bibinfo {author}
  {\bibfnamefont {E.}~\bibnamefont {Ye}}, \bibinfo {author} {\bibfnamefont
  {A.~J.}\ \bibnamefont {Minnich}}, \bibinfo {author} {\bibfnamefont {F.~G.}\
  \bibnamefont {Brandao}},\ and\ \bibinfo {author} {\bibfnamefont {G.~K.-L.}\
  \bibnamefont {Chan}},\ }\bibfield  {title} {\bibinfo {title} {Determining
  eigenstates and thermal states on a quantum computer using quantum imaginary
  time evolution},\ }\href
  {https://doi.org/https://doi.org/10.1038/s41567-019-0704-4} {\bibfield
  {journal} {\bibinfo  {journal} {Nature Physics}\ }\textbf {\bibinfo {volume}
  {16}},\ \bibinfo {pages} {205} (\bibinfo {year} {2020})}\BibitemShut
  {NoStop}%
\bibitem [{\citenamefont {Yeter-Aydeniz}\ \emph {et~al.}(2020)\citenamefont
  {Yeter-Aydeniz}, \citenamefont {Pooser},\ and\ \citenamefont
  {Siopsis}}]{yeter2020practical}%
  \BibitemOpen
  \bibfield  {author} {\bibinfo {author} {\bibfnamefont {K.}~\bibnamefont
  {Yeter-Aydeniz}}, \bibinfo {author} {\bibfnamefont {R.~C.}\ \bibnamefont
  {Pooser}},\ and\ \bibinfo {author} {\bibfnamefont {G.}~\bibnamefont
  {Siopsis}},\ }\bibfield  {title} {\bibinfo {title} {Practical quantum
  computation of chemical and nuclear energy levels using quantum imaginary
  time evolution and lanczos algorithms},\ }\href
  {https://doi.org/https://doi.org/10.1038/s41534-020-00290-1} {\bibfield
  {journal} {\bibinfo  {journal} {npj Quantum Information}\ }\textbf {\bibinfo
  {volume} {6}},\ \bibinfo {pages} {63} (\bibinfo {year} {2020})}\BibitemShut
  {NoStop}%
\bibitem [{\citenamefont {Gomes}\ \emph {et~al.}(2020)\citenamefont {Gomes},
  \citenamefont {Zhang}, \citenamefont {Berthusen}, \citenamefont {Wang},
  \citenamefont {Ho}, \citenamefont {Orth},\ and\ \citenamefont
  {Yao}}]{gomes2020efficient}%
  \BibitemOpen
  \bibfield  {author} {\bibinfo {author} {\bibfnamefont {N.}~\bibnamefont
  {Gomes}}, \bibinfo {author} {\bibfnamefont {F.}~\bibnamefont {Zhang}},
  \bibinfo {author} {\bibfnamefont {N.~F.}\ \bibnamefont {Berthusen}}, \bibinfo
  {author} {\bibfnamefont {C.-Z.}\ \bibnamefont {Wang}}, \bibinfo {author}
  {\bibfnamefont {K.-M.}\ \bibnamefont {Ho}}, \bibinfo {author} {\bibfnamefont
  {P.~P.}\ \bibnamefont {Orth}},\ and\ \bibinfo {author} {\bibfnamefont
  {Y.}~\bibnamefont {Yao}},\ }\bibfield  {title} {\bibinfo {title} {Efficient
  step-merged quantum imaginary time evolution algorithm for quantum
  chemistry},\ }\href
  {https://doi.org/https://doi.org/10.1021/acs.jctc.0c00666} {\bibfield
  {journal} {\bibinfo  {journal} {Journal of Chemical Theory and Computation}\
  }\textbf {\bibinfo {volume} {16}},\ \bibinfo {pages} {6256} (\bibinfo {year}
  {2020})}\BibitemShut {NoStop}%
\bibitem [{\citenamefont {Sun}\ \emph {et~al.}(2021)\citenamefont {Sun},
  \citenamefont {Motta}, \citenamefont {Tazhigulov}, \citenamefont {Tan},
  \citenamefont {Chan},\ and\ \citenamefont {Minnich}}]{sun2021quantum}%
  \BibitemOpen
  \bibfield  {author} {\bibinfo {author} {\bibfnamefont {S.-N.}\ \bibnamefont
  {Sun}}, \bibinfo {author} {\bibfnamefont {M.}~\bibnamefont {Motta}}, \bibinfo
  {author} {\bibfnamefont {R.~N.}\ \bibnamefont {Tazhigulov}}, \bibinfo
  {author} {\bibfnamefont {A.~T.}\ \bibnamefont {Tan}}, \bibinfo {author}
  {\bibfnamefont {G.~K.-L.}\ \bibnamefont {Chan}},\ and\ \bibinfo {author}
  {\bibfnamefont {A.~J.}\ \bibnamefont {Minnich}},\ }\bibfield  {title}
  {\bibinfo {title} {Quantum computation of finite-temperature static and
  dynamical properties of spin systems using quantum imaginary time
  evolution},\ }\href
  {https://doi.org/https://doi.org/10.1103/PRXQuantum.2.010317} {\bibfield
  {journal} {\bibinfo  {journal} {PRX Quantum}\ }\textbf {\bibinfo {volume}
  {2}},\ \bibinfo {pages} {010317} (\bibinfo {year} {2021})}\BibitemShut
  {NoStop}%
\bibitem [{\citenamefont {Lin}\ \emph {et~al.}(2021)\citenamefont {Lin},
  \citenamefont {Dilip}, \citenamefont {Green}, \citenamefont {Smith},\ and\
  \citenamefont {Pollmann}}]{lin2021real}%
  \BibitemOpen
  \bibfield  {author} {\bibinfo {author} {\bibfnamefont {S.-H.}\ \bibnamefont
  {Lin}}, \bibinfo {author} {\bibfnamefont {R.}~\bibnamefont {Dilip}}, \bibinfo
  {author} {\bibfnamefont {A.~G.}\ \bibnamefont {Green}}, \bibinfo {author}
  {\bibfnamefont {A.}~\bibnamefont {Smith}},\ and\ \bibinfo {author}
  {\bibfnamefont {F.}~\bibnamefont {Pollmann}},\ }\bibfield  {title} {\bibinfo
  {title} {Real-and imaginary-time evolution with compressed quantum
  circuits},\ }\href
  {https://doi.org/https://doi.org/10.1103/PRXQuantum.2.010342} {\bibfield
  {journal} {\bibinfo  {journal} {PRX Quantum}\ }\textbf {\bibinfo {volume}
  {2}},\ \bibinfo {pages} {010342} (\bibinfo {year} {2021})}\BibitemShut
  {NoStop}%
\bibitem [{\citenamefont {Kazuki}\ \emph {et~al.}(2021)\citenamefont {Kazuki},
  \citenamefont {Kosugi},\ and\ \citenamefont
  {Matsushita}}]{nishi2021implementation}%
  \BibitemOpen
  \bibfield  {author} {\bibinfo {author} {\bibfnamefont {H.}~\bibnamefont
  {Kazuki}}, \bibinfo {author} {\bibfnamefont {T.}~\bibnamefont {Kosugi}},\
  and\ \bibinfo {author} {\bibfnamefont {Y.-i.}\ \bibnamefont {Matsushita}},\
  }\bibfield  {title} {\bibinfo {title} {Implementation of quantum
  imaginary-time evolution method on nisq devices by introducing nonlocal
  approximation},\ }\href
  {https://doi.org/https://doi.org/10.1038/s41534-021-00409-y} {\bibfield
  {journal} {\bibinfo  {journal} {npj Quantum Information}\ }\textbf {\bibinfo
  {volume} {7}},\ \bibinfo {pages} {85} (\bibinfo {year} {2021})}\BibitemShut
  {NoStop}%
\bibitem [{\citenamefont {Yeter-Aydeniz}\ \emph {et~al.}(2022)\citenamefont
  {Yeter-Aydeniz}, \citenamefont {Moschandreou},\ and\ \citenamefont
  {Siopsis}}]{yeter2022quantum}%
  \BibitemOpen
  \bibfield  {author} {\bibinfo {author} {\bibfnamefont {K.}~\bibnamefont
  {Yeter-Aydeniz}}, \bibinfo {author} {\bibfnamefont {E.}~\bibnamefont
  {Moschandreou}},\ and\ \bibinfo {author} {\bibfnamefont {G.}~\bibnamefont
  {Siopsis}},\ }\bibfield  {title} {\bibinfo {title} {Quantum imaginary-time
  evolution algorithm for quantum field theories with continuous variables},\
  }\href {https://doi.org/https://doi.org/10.1103/PhysRevA.105.012412}
  {\bibfield  {journal} {\bibinfo  {journal} {Physical Review A}\ }\textbf
  {\bibinfo {volume} {105}},\ \bibinfo {pages} {012412} (\bibinfo {year}
  {2022})}\BibitemShut {NoStop}%
\bibitem [{\citenamefont {Tsuchimochi}\ \emph {et~al.}(2023)\citenamefont
  {Tsuchimochi}, \citenamefont {Ryo}, \citenamefont {Ten-No},\ and\
  \citenamefont {Sasasako}}]{tsuchimochi2023improved}%
  \BibitemOpen
  \bibfield  {author} {\bibinfo {author} {\bibfnamefont {T.}~\bibnamefont
  {Tsuchimochi}}, \bibinfo {author} {\bibfnamefont {Y.}~\bibnamefont {Ryo}},
  \bibinfo {author} {\bibfnamefont {S.~L.}\ \bibnamefont {Ten-No}},\ and\
  \bibinfo {author} {\bibfnamefont {K.}~\bibnamefont {Sasasako}},\ }\bibfield
  {title} {\bibinfo {title} {Improved algorithms of quantum imaginary time
  evolution for ground and excited states of molecular systems},\ }\href
  {https://doi.org/https://doi.org/10.1021/acs.jctc.2c00906} {\bibfield
  {journal} {\bibinfo  {journal} {Journal of Chemical Theory and Computation}\
  }\textbf {\bibinfo {volume} {19}},\ \bibinfo {pages} {503} (\bibinfo {year}
  {2023})}\BibitemShut {NoStop}%
\bibitem [{\citenamefont {Liu}\ \emph {et~al.}(2021)\citenamefont {Liu},
  \citenamefont {Liu},\ and\ \citenamefont {Fan}}]{liu2021probabilistic}%
  \BibitemOpen
  \bibfield  {author} {\bibinfo {author} {\bibfnamefont {T.}~\bibnamefont
  {Liu}}, \bibinfo {author} {\bibfnamefont {J.-G.}\ \bibnamefont {Liu}},\ and\
  \bibinfo {author} {\bibfnamefont {H.}~\bibnamefont {Fan}},\ }\bibfield
  {title} {\bibinfo {title} {Probabilistic nonunitary gate in imaginary time
  evolution},\ }\href
  {https://doi.org/https://doi.org/10.1007/s11128-021-03145-6} {\bibfield
  {journal} {\bibinfo  {journal} {Quantum Information Processing}\ }\textbf
  {\bibinfo {volume} {20}},\ \bibinfo {pages} {204} (\bibinfo {year}
  {2021})}\BibitemShut {NoStop}%
\bibitem [{\citenamefont {Kosugi}\ \emph {et~al.}(2022)\citenamefont {Kosugi},
  \citenamefont {Nishiya}, \citenamefont {Nishi},\ and\ \citenamefont
  {Matsushita}}]{kosugi2022imaginary}%
  \BibitemOpen
  \bibfield  {author} {\bibinfo {author} {\bibfnamefont {T.}~\bibnamefont
  {Kosugi}}, \bibinfo {author} {\bibfnamefont {Y.}~\bibnamefont {Nishiya}},
  \bibinfo {author} {\bibfnamefont {H.}~\bibnamefont {Nishi}},\ and\ \bibinfo
  {author} {\bibfnamefont {Y.-i.}\ \bibnamefont {Matsushita}},\ }\bibfield
  {title} {\bibinfo {title} {Imaginary-time evolution using forward and
  backward real-time evolution with a single ancilla: First-quantized
  eigensolver algorithm for quantum chemistry},\ }\href
  {https://doi.org/https://doi.org/10.1103/PhysRevResearch.4.033121} {\bibfield
   {journal} {\bibinfo  {journal} {Physical Review Research}\ }\textbf
  {\bibinfo {volume} {4}},\ \bibinfo {pages} {033121} (\bibinfo {year}
  {2022})}\BibitemShut {NoStop}%
\bibitem [{\citenamefont {Leamer}\ \emph {et~al.}(2024)\citenamefont {Leamer},
  \citenamefont {Bondar},\ and\ \citenamefont {McCaul}}]{leamer2024quantum}%
  \BibitemOpen
  \bibfield  {author} {\bibinfo {author} {\bibfnamefont {J.~M.}\ \bibnamefont
  {Leamer}}, \bibinfo {author} {\bibfnamefont {D.~I.}\ \bibnamefont {Bondar}},\
  and\ \bibinfo {author} {\bibfnamefont {G.}~\bibnamefont {McCaul}},\
  }\bibfield  {title} {\bibinfo {title} {Quantum dynamical emulation},\ }\href
  {https://arxiv.org/abs/2403.03350} {\bibfield  {journal} {\bibinfo  {journal}
  {arXiv preprint arXiv:2403.03350}\ } (\bibinfo {year} {2024})}\BibitemShut
  {NoStop}%
\bibitem [{\citenamefont {McArdle}\ \emph {et~al.}(2019)\citenamefont
  {McArdle}, \citenamefont {Jones}, \citenamefont {Endo}, \citenamefont {Li},
  \citenamefont {Benjamin},\ and\ \citenamefont
  {Yuan}}]{mcardle2019variational}%
  \BibitemOpen
  \bibfield  {author} {\bibinfo {author} {\bibfnamefont {S.}~\bibnamefont
  {McArdle}}, \bibinfo {author} {\bibfnamefont {T.}~\bibnamefont {Jones}},
  \bibinfo {author} {\bibfnamefont {S.}~\bibnamefont {Endo}}, \bibinfo {author}
  {\bibfnamefont {Y.}~\bibnamefont {Li}}, \bibinfo {author} {\bibfnamefont
  {S.~C.}\ \bibnamefont {Benjamin}},\ and\ \bibinfo {author} {\bibfnamefont
  {X.}~\bibnamefont {Yuan}},\ }\bibfield  {title} {\bibinfo {title}
  {Variational ansatz-based quantum simulation of imaginary time evolution},\
  }\href {https://doi.org/https://doi.org/10.1038/s41534-019-0187-2} {\bibfield
   {journal} {\bibinfo  {journal} {npj Quantum Information}\ }\textbf {\bibinfo
  {volume} {5}},\ \bibinfo {pages} {75} (\bibinfo {year} {2019})}\BibitemShut
  {NoStop}%
\bibitem [{\citenamefont {Jones}\ \emph {et~al.}(2019)\citenamefont {Jones},
  \citenamefont {Endo}, \citenamefont {McArdle}, \citenamefont {Yuan},\ and\
  \citenamefont {Benjamin}}]{jones2019variational}%
  \BibitemOpen
  \bibfield  {author} {\bibinfo {author} {\bibfnamefont {T.}~\bibnamefont
  {Jones}}, \bibinfo {author} {\bibfnamefont {S.}~\bibnamefont {Endo}},
  \bibinfo {author} {\bibfnamefont {S.}~\bibnamefont {McArdle}}, \bibinfo
  {author} {\bibfnamefont {X.}~\bibnamefont {Yuan}},\ and\ \bibinfo {author}
  {\bibfnamefont {S.~C.}\ \bibnamefont {Benjamin}},\ }\bibfield  {title}
  {\bibinfo {title} {Variational quantum algorithms for discovering hamiltonian
  spectra},\ }\href
  {https://doi.org/https://doi.org/10.1103/PhysRevA.99.062304} {\bibfield
  {journal} {\bibinfo  {journal} {Physical Review A}\ }\textbf {\bibinfo
  {volume} {99}},\ \bibinfo {pages} {062304} (\bibinfo {year}
  {2019})}\BibitemShut {NoStop}%
\bibitem [{\citenamefont {Helmke}\ and\ \citenamefont
  {Moore}(2012)}]{helmke2012optimization}%
  \BibitemOpen
  \bibfield  {author} {\bibinfo {author} {\bibfnamefont {U.}~\bibnamefont
  {Helmke}}\ and\ \bibinfo {author} {\bibfnamefont {J.~B.}\ \bibnamefont
  {Moore}},\ }\href@noop {} {\emph {\bibinfo {title} {Optimization and
  dynamical systems}}}\ (\bibinfo  {publisher} {Springer Science \& Business
  Media},\ \bibinfo {year} {2012})\BibitemShut {NoStop}%
\bibitem [{\citenamefont {Schulte-Herbr{\"u}ggen}\ \emph
  {et~al.}(2010)\citenamefont {Schulte-Herbr{\"u}ggen}, \citenamefont {Glaser},
  \citenamefont {Dirr},\ and\ \citenamefont {Helmke}}]{schulte2010gradient}%
  \BibitemOpen
  \bibfield  {author} {\bibinfo {author} {\bibfnamefont {T.}~\bibnamefont
  {Schulte-Herbr{\"u}ggen}}, \bibinfo {author} {\bibfnamefont {S.~J.}\
  \bibnamefont {Glaser}}, \bibinfo {author} {\bibfnamefont {G.}~\bibnamefont
  {Dirr}},\ and\ \bibinfo {author} {\bibfnamefont {U.}~\bibnamefont {Helmke}},\
  }\bibfield  {title} {\bibinfo {title} {Gradient flows for optimization in
  quantum information and quantum dynamics: foundations and applications},\
  }\href {https://doi.org/https://doi.org/10.1142/S0129055X10004053} {\bibfield
   {journal} {\bibinfo  {journal} {Reviews in Mathematical Physics}\ }\textbf
  {\bibinfo {volume} {22}},\ \bibinfo {pages} {597} (\bibinfo {year}
  {2010})}\BibitemShut {NoStop}%
\bibitem [{\citenamefont {Malvetti}\ \emph {et~al.}(2024)\citenamefont
  {Malvetti}, \citenamefont {Arenz}, \citenamefont {Dirr},\ and\ \citenamefont
  {Schulte-Herbr{\"u}ggen}}]{malvetti2024randomized}%
  \BibitemOpen
  \bibfield  {author} {\bibinfo {author} {\bibfnamefont {E.}~\bibnamefont
  {Malvetti}}, \bibinfo {author} {\bibfnamefont {C.}~\bibnamefont {Arenz}},
  \bibinfo {author} {\bibfnamefont {G.}~\bibnamefont {Dirr}},\ and\ \bibinfo
  {author} {\bibfnamefont {T.}~\bibnamefont {Schulte-Herbr{\"u}ggen}},\
  }\bibfield  {title} {\bibinfo {title} {Randomized gradient descents on
  riemannian manifolds: Almost sure convergence to global minima in and beyond
  quantum optimization},\ }\href {https://arxiv.org/abs/2405.12039} {\bibfield
  {journal} {\bibinfo  {journal} {arXiv preprint arXiv:2405.12039}\ } (\bibinfo
  {year} {2024})}\BibitemShut {NoStop}%
\bibitem [{\citenamefont {Wiersema}\ and\ \citenamefont
  {Killoran}(2023)}]{wiersema2023optimizing}%
  \BibitemOpen
  \bibfield  {author} {\bibinfo {author} {\bibfnamefont {R.}~\bibnamefont
  {Wiersema}}\ and\ \bibinfo {author} {\bibfnamefont {N.}~\bibnamefont
  {Killoran}},\ }\bibfield  {title} {\bibinfo {title} {Optimizing quantum
  circuits with riemannian gradient flow},\ }\href
  {https://doi.org/https://doi.org/10.1103/PhysRevA.107.062421} {\bibfield
  {journal} {\bibinfo  {journal} {Physical Review A}\ }\textbf {\bibinfo
  {volume} {107}},\ \bibinfo {pages} {062421} (\bibinfo {year}
  {2023})}\BibitemShut {NoStop}%
\bibitem [{\citenamefont {Magann}\ \emph {et~al.}(2023)\citenamefont {Magann},
  \citenamefont {Economou},\ and\ \citenamefont
  {Arenz}}]{magann2023randomized}%
  \BibitemOpen
  \bibfield  {author} {\bibinfo {author} {\bibfnamefont {A.~B.}\ \bibnamefont
  {Magann}}, \bibinfo {author} {\bibfnamefont {S.~E.}\ \bibnamefont
  {Economou}},\ and\ \bibinfo {author} {\bibfnamefont {C.}~\bibnamefont
  {Arenz}},\ }\bibfield  {title} {\bibinfo {title} {Randomized adaptive quantum
  state preparation},\ }\href
  {https://doi.org/https://doi.org/10.1103/PhysRevResearch.5.033227} {\bibfield
   {journal} {\bibinfo  {journal} {Physical Review Research}\ }\textbf
  {\bibinfo {volume} {5}},\ \bibinfo {pages} {033227} (\bibinfo {year}
  {2023})}\BibitemShut {NoStop}%
\bibitem [{\citenamefont {Grimsley}\ \emph {et~al.}(2019)\citenamefont
  {Grimsley}, \citenamefont {Economou}, \citenamefont {Barnes},\ and\
  \citenamefont {Mayhall}}]{grimsley2019adaptive}%
  \BibitemOpen
  \bibfield  {author} {\bibinfo {author} {\bibfnamefont {H.~R.}\ \bibnamefont
  {Grimsley}}, \bibinfo {author} {\bibfnamefont {S.~E.}\ \bibnamefont
  {Economou}}, \bibinfo {author} {\bibfnamefont {E.}~\bibnamefont {Barnes}},\
  and\ \bibinfo {author} {\bibfnamefont {N.~J.}\ \bibnamefont {Mayhall}},\
  }\bibfield  {title} {\bibinfo {title} {An adaptive variational algorithm for
  exact molecular simulations on a quantum computer},\ }\href
  {https://doi.org/https://doi.org/10.1038/s41467-019-10988-2} {\bibfield
  {journal} {\bibinfo  {journal} {Nature communications}\ }\textbf {\bibinfo
  {volume} {10}},\ \bibinfo {pages} {3007} (\bibinfo {year}
  {2019})}\BibitemShut {NoStop}%
\bibitem [{\citenamefont {Tang}\ \emph {et~al.}(2021)\citenamefont {Tang},
  \citenamefont {Shkolnikov}, \citenamefont {Barron}, \citenamefont {Grimsley},
  \citenamefont {Mayhall}, \citenamefont {Barnes},\ and\ \citenamefont
  {Economou}}]{tang2021qubit}%
  \BibitemOpen
  \bibfield  {author} {\bibinfo {author} {\bibfnamefont {H.~L.}\ \bibnamefont
  {Tang}}, \bibinfo {author} {\bibfnamefont {V.}~\bibnamefont {Shkolnikov}},
  \bibinfo {author} {\bibfnamefont {G.~S.}\ \bibnamefont {Barron}}, \bibinfo
  {author} {\bibfnamefont {H.~R.}\ \bibnamefont {Grimsley}}, \bibinfo {author}
  {\bibfnamefont {N.~J.}\ \bibnamefont {Mayhall}}, \bibinfo {author}
  {\bibfnamefont {E.}~\bibnamefont {Barnes}},\ and\ \bibinfo {author}
  {\bibfnamefont {S.~E.}\ \bibnamefont {Economou}},\ }\bibfield  {title}
  {\bibinfo {title} {qubit-adapt-vqe: An adaptive algorithm for constructing
  hardware-efficient ans{\"a}tze on a quantum processor},\ }\href
  {https://doi.org/https://doi.org/10.1103/PRXQuantum.2.020310} {\bibfield
  {journal} {\bibinfo  {journal} {PRX Quantum}\ }\textbf {\bibinfo {volume}
  {2}},\ \bibinfo {pages} {020310} (\bibinfo {year} {2021})}\BibitemShut
  {NoStop}%
\bibitem [{\citenamefont {Magann}\ \emph
  {et~al.}(2022{\natexlab{a}})\citenamefont {Magann}, \citenamefont {Rudinger},
  \citenamefont {Grace},\ and\ \citenamefont {Sarovar}}]{magann2022feedback}%
  \BibitemOpen
  \bibfield  {author} {\bibinfo {author} {\bibfnamefont {A.~B.}\ \bibnamefont
  {Magann}}, \bibinfo {author} {\bibfnamefont {K.~M.}\ \bibnamefont
  {Rudinger}}, \bibinfo {author} {\bibfnamefont {M.~D.}\ \bibnamefont
  {Grace}},\ and\ \bibinfo {author} {\bibfnamefont {M.}~\bibnamefont
  {Sarovar}},\ }\bibfield  {title} {\bibinfo {title} {Feedback-based quantum
  optimization},\ }\href
  {https://doi.org/https://doi.org/10.1103/PhysRevLett.129.250502} {\bibfield
  {journal} {\bibinfo  {journal} {Physical Review Letters}\ }\textbf {\bibinfo
  {volume} {129}},\ \bibinfo {pages} {250502} (\bibinfo {year}
  {2022}{\natexlab{a}})}\BibitemShut {NoStop}%
\bibitem [{\citenamefont {Larsen}\ \emph {et~al.}(2024)\citenamefont {Larsen},
  \citenamefont {Grace}, \citenamefont {Baczewski},\ and\ \citenamefont
  {Magann}}]{larsen2024feedback}%
  \BibitemOpen
  \bibfield  {author} {\bibinfo {author} {\bibfnamefont {J.~B.}\ \bibnamefont
  {Larsen}}, \bibinfo {author} {\bibfnamefont {M.~D.}\ \bibnamefont {Grace}},
  \bibinfo {author} {\bibfnamefont {A.~D.}\ \bibnamefont {Baczewski}},\ and\
  \bibinfo {author} {\bibfnamefont {A.~B.}\ \bibnamefont {Magann}},\ }\bibfield
   {title} {\bibinfo {title} {Feedback-based quantum algorithms for ground
  state preparation},\ }\href
  {https://doi.org/https://doi.org/10.1103/PhysRevResearch.6.033336} {\bibfield
   {journal} {\bibinfo  {journal} {Physical Review Research}\ }\textbf
  {\bibinfo {volume} {6}},\ \bibinfo {pages} {033336} (\bibinfo {year}
  {2024})}\BibitemShut {NoStop}%
\bibitem [{\citenamefont {Magann}\ \emph
  {et~al.}(2022{\natexlab{b}})\citenamefont {Magann}, \citenamefont {Rudinger},
  \citenamefont {Grace},\ and\ \citenamefont {Sarovar}}]{magann2022lyapunov}%
  \BibitemOpen
  \bibfield  {author} {\bibinfo {author} {\bibfnamefont {A.~B.}\ \bibnamefont
  {Magann}}, \bibinfo {author} {\bibfnamefont {K.~M.}\ \bibnamefont
  {Rudinger}}, \bibinfo {author} {\bibfnamefont {M.~D.}\ \bibnamefont
  {Grace}},\ and\ \bibinfo {author} {\bibfnamefont {M.}~\bibnamefont
  {Sarovar}},\ }\bibfield  {title} {\bibinfo {title} {Lyapunov-control-inspired
  strategies for quantum combinatorial optimization},\ }\href
  {https://doi.org/https://doi.org/10.1103/PhysRevA.106.062414} {\bibfield
  {journal} {\bibinfo  {journal} {Physical Review A}\ }\textbf {\bibinfo
  {volume} {106}},\ \bibinfo {pages} {062414} (\bibinfo {year}
  {2022}{\natexlab{b}})}\BibitemShut {NoStop}%
\bibitem [{\citenamefont {Tang}\ \emph {et~al.}(2024)\citenamefont {Tang},
  \citenamefont {Chen}, \citenamefont {Biswas}, \citenamefont {Magann},
  \citenamefont {Arenz},\ and\ \citenamefont {Economou}}]{tang2024non}%
  \BibitemOpen
  \bibfield  {author} {\bibinfo {author} {\bibfnamefont {H.~L.}\ \bibnamefont
  {Tang}}, \bibinfo {author} {\bibfnamefont {Y.}~\bibnamefont {Chen}}, \bibinfo
  {author} {\bibfnamefont {P.}~\bibnamefont {Biswas}}, \bibinfo {author}
  {\bibfnamefont {A.~B.}\ \bibnamefont {Magann}}, \bibinfo {author}
  {\bibfnamefont {C.}~\bibnamefont {Arenz}},\ and\ \bibinfo {author}
  {\bibfnamefont {S.~E.}\ \bibnamefont {Economou}},\ }\bibfield  {title}
  {\bibinfo {title} {Non-variational adapt algorithm for quantum simulations},\
  }\href {https://arxiv.org/abs/2411.09736} {\bibfield  {journal} {\bibinfo
  {journal} {arXiv preprint arXiv:2411.09736}\ } (\bibinfo {year}
  {2024})}\BibitemShut {NoStop}%
\bibitem [{\citenamefont {Stokes}\ \emph {et~al.}(2023)\citenamefont {Stokes},
  \citenamefont {Chen},\ and\ \citenamefont
  {Veerapaneni}}]{stokes2023numerical}%
  \BibitemOpen
  \bibfield  {author} {\bibinfo {author} {\bibfnamefont {J.}~\bibnamefont
  {Stokes}}, \bibinfo {author} {\bibfnamefont {B.}~\bibnamefont {Chen}},\ and\
  \bibinfo {author} {\bibfnamefont {S.}~\bibnamefont {Veerapaneni}},\
  }\bibfield  {title} {\bibinfo {title} {Numerical and geometrical aspects of
  flow-based variational quantum monte carlo},\ }\href
  {https://doi.org/https://dx.doi.org/10.1088/2632-2153/acc8b9} {\bibfield
  {journal} {\bibinfo  {journal} {Machine Learning: Science and Technology}\
  }\textbf {\bibinfo {volume} {4}},\ \bibinfo {pages} {021001} (\bibinfo {year}
  {2023})}\BibitemShut {NoStop}%
\bibitem [{\citenamefont {Gluza}\ \emph {et~al.}(2024)\citenamefont {Gluza},
  \citenamefont {Son}, \citenamefont {Tiang}, \citenamefont {Suzuki},
  \citenamefont {Holmes},\ and\ \citenamefont {Ng}}]{gluza2024double}%
  \BibitemOpen
  \bibfield  {author} {\bibinfo {author} {\bibfnamefont {M.}~\bibnamefont
  {Gluza}}, \bibinfo {author} {\bibfnamefont {J.}~\bibnamefont {Son}}, \bibinfo
  {author} {\bibfnamefont {B.~H.}\ \bibnamefont {Tiang}}, \bibinfo {author}
  {\bibfnamefont {Y.}~\bibnamefont {Suzuki}}, \bibinfo {author} {\bibfnamefont
  {Z.}~\bibnamefont {Holmes}},\ and\ \bibinfo {author} {\bibfnamefont {N.~H.}\
  \bibnamefont {Ng}},\ }\bibfield  {title} {\bibinfo {title} {Double-bracket
  quantum algorithms for quantum imaginary-time evolution},\ }\href
  {https://arxiv.org/abs/2412.04554} {\bibfield  {journal} {\bibinfo  {journal}
  {arXiv preprint arXiv:2412.04554}\ } (\bibinfo {year} {2024})}\BibitemShut
  {NoStop}%
\bibitem [{\citenamefont {Gluza}(2024)}]{gluza2024double_2}%
  \BibitemOpen
  \bibfield  {author} {\bibinfo {author} {\bibfnamefont {M.}~\bibnamefont
  {Gluza}},\ }\bibfield  {title} {\bibinfo {title} {Double-bracket quantum
  algorithms for diagonalization},\ }\href
  {https://doi.org/https://doi.org/10.22331/q-2024-04-09-1316} {\bibfield
  {journal} {\bibinfo  {journal} {Quantum}\ }\textbf {\bibinfo {volume} {8}},\
  \bibinfo {pages} {1316} (\bibinfo {year} {2024})}\BibitemShut {NoStop}%
\bibitem [{\citenamefont {Brockett}(1989)}]{brockett1989least}%
  \BibitemOpen
  \bibfield  {author} {\bibinfo {author} {\bibfnamefont {R.~W.}\ \bibnamefont
  {Brockett}},\ }\bibfield  {title} {\bibinfo {title} {Least squares matching
  problems},\ }\href
  {https://doi.org/https://doi.org/10.1016/0024-3795(89)90675-7} {\bibfield
  {journal} {\bibinfo  {journal} {Linear Algebra and its applications}\
  }\textbf {\bibinfo {volume} {122}},\ \bibinfo {pages} {761} (\bibinfo {year}
  {1989})}\BibitemShut {NoStop}%
\bibitem [{\citenamefont {Brockett}(1993)}]{brockett1993differential}%
  \BibitemOpen
  \bibfield  {author} {\bibinfo {author} {\bibfnamefont {R.~W.}\ \bibnamefont
  {Brockett}},\ }\bibfield  {title} {\bibinfo {title} {Differential geometry
  and the design of gradient algorithms},\ }in\ \href
  {https://doi.org/https://cir.nii.ac.jp/crid/1572543024700578688} {\emph
  {\bibinfo {booktitle} {Proc. Symp. Pure Math., AMS}}},\ Vol.~\bibinfo
  {volume} {54}\ (\bibinfo {year} {1993})\ pp.\ \bibinfo {pages}
  {69--92}\BibitemShut {NoStop}%
\bibitem [{\citenamefont {Dawson}\ \emph {et~al.}(2008)\citenamefont {Dawson},
  \citenamefont {Eisert},\ and\ \citenamefont {Osborne}}]{dawson2008unifying}%
  \BibitemOpen
  \bibfield  {author} {\bibinfo {author} {\bibfnamefont {C.~M.}\ \bibnamefont
  {Dawson}}, \bibinfo {author} {\bibfnamefont {J.}~\bibnamefont {Eisert}},\
  and\ \bibinfo {author} {\bibfnamefont {T.~J.}\ \bibnamefont {Osborne}},\
  }\bibfield  {title} {\bibinfo {title} {Unifying variational methods for
  simulating quantum many-body systems},\ }\href
  {https://doi.org/https://doi.org/10.1103/PhysRevLett.100.130501} {\bibfield
  {journal} {\bibinfo  {journal} {Physical review letters}\ }\textbf {\bibinfo
  {volume} {100}},\ \bibinfo {pages} {130501} (\bibinfo {year}
  {2008})}\BibitemShut {NoStop}%
\bibitem [{\citenamefont {Hastings}(2022)}]{hastings2022lieb}%
  \BibitemOpen
  \bibfield  {author} {\bibinfo {author} {\bibfnamefont {M.~B.}\ \bibnamefont
  {Hastings}},\ }\bibfield  {title} {\bibinfo {title} {On lieb--robinson bounds
  for the double bracket flow},\ }\href
  {https://doi.org/https://doi.org/10.4171/90-1} {\bibfield  {journal}
  {\bibinfo  {journal} {The Physics and Mathematics of Elliott Lieb}\ ,\
  \bibinfo {pages} {515}} (\bibinfo {year} {2022})}\BibitemShut {NoStop}%
\bibitem [{\citenamefont {Cerezo}\ \emph {et~al.}(2021)\citenamefont {Cerezo},
  \citenamefont {Arrasmith}, \citenamefont {Babbush}, \citenamefont {Benjamin},
  \citenamefont {Endo}, \citenamefont {Fujii}, \citenamefont {McClean},
  \citenamefont {Mitarai}, \citenamefont {Yuan}, \citenamefont {Cincio} \emph
  {et~al.}}]{cerezo2021variational}%
  \BibitemOpen
  \bibfield  {author} {\bibinfo {author} {\bibfnamefont {M.}~\bibnamefont
  {Cerezo}}, \bibinfo {author} {\bibfnamefont {A.}~\bibnamefont {Arrasmith}},
  \bibinfo {author} {\bibfnamefont {R.}~\bibnamefont {Babbush}}, \bibinfo
  {author} {\bibfnamefont {S.~C.}\ \bibnamefont {Benjamin}}, \bibinfo {author}
  {\bibfnamefont {S.}~\bibnamefont {Endo}}, \bibinfo {author} {\bibfnamefont
  {K.}~\bibnamefont {Fujii}}, \bibinfo {author} {\bibfnamefont {J.~R.}\
  \bibnamefont {McClean}}, \bibinfo {author} {\bibfnamefont {K.}~\bibnamefont
  {Mitarai}}, \bibinfo {author} {\bibfnamefont {X.}~\bibnamefont {Yuan}},
  \bibinfo {author} {\bibfnamefont {L.}~\bibnamefont {Cincio}}, \emph
  {et~al.},\ }\bibfield  {title} {\bibinfo {title} {Variational quantum
  algorithms},\ }\href
  {https://doi.org/https://doi.org/10.1038/s42254-021-00348-9} {\bibfield
  {journal} {\bibinfo  {journal} {Nature Reviews Physics}\ }\textbf {\bibinfo
  {volume} {3}},\ \bibinfo {pages} {625} (\bibinfo {year} {2021})}\BibitemShut
  {NoStop}%
\bibitem [{Note1()}]{Note1}%
  \BibitemOpen
  \bibinfo {note} {With a slight abuse of notation, from now on we denote the
  Riemannian gradient by $\protect \text {grad}J= [H,\protect \ensuremath
  {\mathinner {|{\phi }\rangle }}\protect \ensuremath {\mathinner {\langle
  {\phi }|}}]$, noting that due to the invariance of the Hilber-Schmidt inner
  product $\langle X,Y\rangle =\protect \text {Tr}\{X^{\dagger }Y\}$ with
  respect to $U$, the inner product between tangent space elements $X,Y\in
  T_{U}\protect \text {SU}(d)$ and the inner product between elements $X,Y\in
  \protect \mathfrak {su}(d)$ is the same}\BibitemShut {NoStop}%
\bibitem [{\citenamefont {White}(2009)}]{white2009minimally}%
  \BibitemOpen
  \bibfield  {author} {\bibinfo {author} {\bibfnamefont {S.~R.}\ \bibnamefont
  {White}},\ }\bibfield  {title} {\bibinfo {title} {Minimally entangled typical
  quantum states at finite temperature},\ }\href
  {https://doi.org/https://doi.org/10.1103/PhysRevLett.102.190601} {\bibfield
  {journal} {\bibinfo  {journal} {Physical review letters}\ }\textbf {\bibinfo
  {volume} {102}},\ \bibinfo {pages} {190601} (\bibinfo {year}
  {2009})}\BibitemShut {NoStop}%
\bibitem [{\citenamefont {Mitarai}\ \emph {et~al.}(2018)\citenamefont
  {Mitarai}, \citenamefont {Negoro}, \citenamefont {Kitagawa},\ and\
  \citenamefont {Fujii}}]{mitarai2018quantum}%
  \BibitemOpen
  \bibfield  {author} {\bibinfo {author} {\bibfnamefont {K.}~\bibnamefont
  {Mitarai}}, \bibinfo {author} {\bibfnamefont {M.}~\bibnamefont {Negoro}},
  \bibinfo {author} {\bibfnamefont {M.}~\bibnamefont {Kitagawa}},\ and\
  \bibinfo {author} {\bibfnamefont {K.}~\bibnamefont {Fujii}},\ }\bibfield
  {title} {\bibinfo {title} {Quantum circuit learning},\ }\href
  {https://doi.org/https://doi.org/10.1103/PhysRevA.98.032309} {\bibfield
  {journal} {\bibinfo  {journal} {Physical Review A}\ }\textbf {\bibinfo
  {volume} {98}},\ \bibinfo {pages} {032309} (\bibinfo {year}
  {2018})}\BibitemShut {NoStop}%
\bibitem [{\citenamefont {Schuld}\ \emph {et~al.}(2019)\citenamefont {Schuld},
  \citenamefont {Bergholm}, \citenamefont {Gogolin}, \citenamefont {Izaac},\
  and\ \citenamefont {Killoran}}]{schuld2019evaluating}%
  \BibitemOpen
  \bibfield  {author} {\bibinfo {author} {\bibfnamefont {M.}~\bibnamefont
  {Schuld}}, \bibinfo {author} {\bibfnamefont {V.}~\bibnamefont {Bergholm}},
  \bibinfo {author} {\bibfnamefont {C.}~\bibnamefont {Gogolin}}, \bibinfo
  {author} {\bibfnamefont {J.}~\bibnamefont {Izaac}},\ and\ \bibinfo {author}
  {\bibfnamefont {N.}~\bibnamefont {Killoran}},\ }\bibfield  {title} {\bibinfo
  {title} {Evaluating analytic gradients on quantum hardware},\ }\href
  {https://doi.org/https://doi.org/10.1103/PhysRevA.99.032331} {\bibfield
  {journal} {\bibinfo  {journal} {Physical Review A}\ }\textbf {\bibinfo
  {volume} {99}},\ \bibinfo {pages} {032331} (\bibinfo {year}
  {2019})}\BibitemShut {NoStop}%
\bibitem [{\citenamefont {Gutman}\ and\ \citenamefont
  {Ho-Nguyen}(2023)}]{gutman2023coordinate}%
  \BibitemOpen
  \bibfield  {author} {\bibinfo {author} {\bibfnamefont {D.~H.}\ \bibnamefont
  {Gutman}}\ and\ \bibinfo {author} {\bibfnamefont {N.}~\bibnamefont
  {Ho-Nguyen}},\ }\bibfield  {title} {\bibinfo {title} {Coordinate descent
  without coordinates: Tangent subspace descent on riemannian manifolds},\
  }\href {https://doi.org/https://doi.org/10.1287/moor.2022.1253} {\bibfield
  {journal} {\bibinfo  {journal} {Mathematics of Operations Research}\ }\textbf
  {\bibinfo {volume} {48}},\ \bibinfo {pages} {127} (\bibinfo {year}
  {2023})}\BibitemShut {NoStop}%
\bibitem [{\citenamefont {Kempe}\ \emph {et~al.}(2006)\citenamefont {Kempe},
  \citenamefont {Kitaev},\ and\ \citenamefont {Regev}}]{kempe2006complexity}%
  \BibitemOpen
  \bibfield  {author} {\bibinfo {author} {\bibfnamefont {J.}~\bibnamefont
  {Kempe}}, \bibinfo {author} {\bibfnamefont {A.}~\bibnamefont {Kitaev}},\ and\
  \bibinfo {author} {\bibfnamefont {O.}~\bibnamefont {Regev}},\ }\bibfield
  {title} {\bibinfo {title} {The complexity of the local hamiltonian problem},\
  }\href {https://doi.org/https://doi.org/10.1137/S0097539704445226} {\bibfield
   {journal} {\bibinfo  {journal} {Siam journal on computing}\ }\textbf
  {\bibinfo {volume} {35}},\ \bibinfo {pages} {1070} (\bibinfo {year}
  {2006})}\BibitemShut {NoStop}%
\bibitem [{\citenamefont {Kong}\ and\ \citenamefont
  {Tao}(2024)}]{kong2024quantitative}%
  \BibitemOpen
  \bibfield  {author} {\bibinfo {author} {\bibfnamefont {L.}~\bibnamefont
  {Kong}}\ and\ \bibinfo {author} {\bibfnamefont {M.}~\bibnamefont {Tao}},\
  }\bibfield  {title} {\bibinfo {title} {Quantitative convergences of lie group
  momentum optimizers},\ }\bibfield  {journal} {\bibinfo  {journal} {arXiv
  preprint arXiv:2405.20390}\ }\href
  {https://doi.org/https://arxiv.org/abs/2405.20390}
  {https://arxiv.org/abs/2405.20390} (\bibinfo {year} {2024})\BibitemShut
  {NoStop}%
\bibitem [{\citenamefont {Vanderstraeten}\ \emph {et~al.}(2019)\citenamefont
  {Vanderstraeten}, \citenamefont {Haegeman},\ and\ \citenamefont
  {Verstraete}}]{vanderstraeten2019tangent}%
  \BibitemOpen
  \bibfield  {author} {\bibinfo {author} {\bibfnamefont {L.}~\bibnamefont
  {Vanderstraeten}}, \bibinfo {author} {\bibfnamefont {J.}~\bibnamefont
  {Haegeman}},\ and\ \bibinfo {author} {\bibfnamefont {F.}~\bibnamefont
  {Verstraete}},\ }\bibfield  {title} {\bibinfo {title} {Tangent-space methods
  for uniform matrix product states},\ }\href
  {https://doi.org/10.21468/SciPostPhysLectNotes.7} {\bibfield  {journal}
  {\bibinfo  {journal} {SciPost Physics Lecture Notes}\ ,\ \bibinfo {pages}
  {007}} (\bibinfo {year} {2019})}\BibitemShut {NoStop}%
\bibitem [{\citenamefont {Miao}\ and\ \citenamefont
  {Barthel}(2023)}]{miao2023quantum}%
  \BibitemOpen
  \bibfield  {author} {\bibinfo {author} {\bibfnamefont {Q.}~\bibnamefont
  {Miao}}\ and\ \bibinfo {author} {\bibfnamefont {T.}~\bibnamefont {Barthel}},\
  }\bibfield  {title} {\bibinfo {title} {Quantum-classical eigensolver using
  multiscale entanglement renormalization},\ }\href
  {https://doi.org/https://doi.org/10.1103/PhysRevResearch.5.033141} {\bibfield
   {journal} {\bibinfo  {journal} {Physical Review Research}\ }\textbf
  {\bibinfo {volume} {5}},\ \bibinfo {pages} {033141} (\bibinfo {year}
  {2023})}\BibitemShut {NoStop}%
\bibitem [{\citenamefont {Dunjko}\ and\ \citenamefont
  {Briegel}(2018)}]{dunjko2018machine}%
  \BibitemOpen
  \bibfield  {author} {\bibinfo {author} {\bibfnamefont {V.}~\bibnamefont
  {Dunjko}}\ and\ \bibinfo {author} {\bibfnamefont {H.~J.}\ \bibnamefont
  {Briegel}},\ }\bibfield  {title} {\bibinfo {title} {Machine learning \&
  artificial intelligence in the quantum domain: a review of recent progress},\
  }\href {https://doi.org/https://dx.doi.org/10.1088/1361-6633/aab406}
  {\bibfield  {journal} {\bibinfo  {journal} {Reports on Progress in Physics}\
  }\textbf {\bibinfo {volume} {81}},\ \bibinfo {pages} {074001} (\bibinfo
  {year} {2018})}\BibitemShut {NoStop}%
\end{thebibliography}%

\onecolumngrid

\newpage
\section*{Appendix}
\subsection{Proof of Lemma 1 and Theorem 1} 
\label{sec:Lemma1Theorem1}
\vspace{-0.25cm}
We define the imaginary time evolved state 
\begin{align}
\ket{\psi(\beta)}=\frac{e^{-\beta H}\ket{\psi_{0}}}{\Vert e^{-\beta H}\ket{\psi_{0}}\Vert },
\end{align}
for some initial state $\ket{\psi_{0}}$ and further define the Riemannian gradient descent (RGD) update via $\ket{\phi_{k}}=U_{\phi_{k-1}}(\Delta \beta)\ket{\phi_{k-1}}$  where
\begin{align}
U_{\phi}(\Delta \beta)=e^{-\Delta \beta [H,\ket{\phi}\bra{\phi}]}
\end{align}
with step size $\Delta \beta=\frac{\beta}{n}$. In contrast to the main body of the manuscript, here we indicate that the RGD update step $U_{\phi}$ depends on the state $\ket{\phi}$ the update is applied to. Throughout this work we assume that the initial state $\ket{\psi_{0}}$ of ITE is the same for RGD. 
\begin{mdframed}
\textbf{Lemma 1}: \textit{Let}
$\ket{\psi(\Delta\beta)}=\frac{e^{-\Delta \beta H}\ket{\psi}}{\Vert e^{-\Delta \beta H}\ket{\psi}\Vert } $ and $ \ket{\phi(\Delta \beta)}=U_{\psi}(\Delta \beta)\ket{\psi}$ 
\textit{be the states created by one step of imaginary time evolution and one step of RGD with step size $\Delta \beta$.  
 Then for any $\ket{\psi}$, }
 \begin{align}
 \Vert \ket{\psi(\Delta \beta)}-\ket{\phi(\Delta \beta)} \Vert \leq  6\Delta\beta^{2}\Vert H\Vert_{\infty}^{2}
 \end{align}
 \end{mdframed}
\textit{Proof:} 
By the triangle inequality the error between the imaginary time evolved state \(\ket{\psi(\Delta\beta)}\) and the state \(\ket{\phi(\Delta \beta)}\) is upper bounded by
\begin{align}
   \Vert \ket{\psi(\Delta \beta)}-\ket{\phi(\Delta \beta)} \Vert &= \left\| \frac{e^{- \Delta \beta H} \ket{\psi}}{\left\| e^{- \Delta \beta H} \ket{\psi} \right\| } - e^{ \Delta \beta [\ket{\psi}\bra{\psi},H]} \ket{\psi} \right\|\\
   &\leq \left\| \left( 1 + \left( -H + \langle \psi | H | \psi \rangle \right)  \Delta \beta \right) | \psi \rangle  - \left( 1 + \left( -H + \langle \psi | H | \psi \rangle \right) \Delta \beta \right)| \psi \rangle  \right\|  + R^{\prime} + R \\
   &= R^{\prime}+ R, 
\end{align}
where the remainder terms $R^{\prime}$ and $R$ for the first order of the Taylor expansion of $\frac{e^{- \Delta \beta H}}{\left\| e^{- \Delta \beta H} \ket{\psi} \right\| } $ and $e^{ \Delta \beta [\ket{\psi}\bra{\psi},H]}$ in $\Delta \beta$ can be upper bounded using the Lagrange form of the remainder. We find  
\begin{align}
    R^{\prime} &\leq  \left\|\left.\Delta \beta^2 \frac{1}{2!} \frac{d^2}{d x^{2}} \frac{e^{- x H}}{\left\| e^{- x H} \ket{\psi} \right\| } \ket{\psi}\right|_{x=\Delta\beta} \right\| 
    \\
    &\leq \left\| \frac{\Delta \beta^{2}}{2}\left[ H^{2} \ket{\psi^{'}} - 2 \langle H \rangle H \ket{\psi^{'}} - 2 \langle H^{2}\rangle \ket{\psi^{'}} +3 \langle H\rangle^{2} \ket{\psi^{'}} \right] \right\| \leq 4 \Delta \beta^2 \Vert H\Vert_{\infty}^{2},
\end{align}
where $\ket{\psi^{'}}$ is the normalised state $\ket{\psi^{'}} = \frac{e^{-xH}\ket{\psi}}{\left\|e^{-xH}\ket{\psi}\right\|}$ at $x= \Delta \beta$, and $\langle H\rangle$ ($\langle H^{2}\rangle$) is the expectation value of $H$ ($H^{2}$) with respect to $\ket{\psi^{'}}$.
Similarly we can bound
\begin{align}
    R \leq  \left\|\left.\Delta \beta^2 \frac{1}{2!} \frac{d^2}{d x^{2}} e^{ \Delta \beta [\ket{\psi}\bra{\psi},H]} \ket{\psi}\right|_{x=\Delta\beta} \right\| \leq 2  \Delta \beta^{2} \Vert H\Vert_{\infty}^{2}.
\end{align}
We thus arrive at 
\begin{align}
   \Vert \ket{\psi(\Delta \beta)}-\ket{\phi(\Delta \beta)} \Vert &\leq 6\Delta \beta^{2}\Vert H\Vert_{\infty}^{2}
\end{align}
which completes the proof of Lemma 1. 
\qed
\newpage
\begin{mdframed}
\textbf{Theorem 1}: \textit{The error $\epsilon_{n}=\Vert\ket{\psi(\beta)}-\ket{\phi_{n}}\Vert $ between the imaginary time evolved state $\ket{\psi(\beta)}$ and the state $\ket{\phi_{n}}$ created by $n$ steps of RGD with step size $\Delta \beta=\frac{\beta}{n}$ is upper bound by}
\begin{align}
\epsilon_{n}\leq \frac{5}{2} \Delta \beta \Vert H\Vert_{\infty}\left(e^{4\beta \Vert H\Vert_{\infty}}-1\right) 
\end{align}
\end{mdframed}
\textit{Proof:} We first note that the imaginary time evolved state $\ket{\psi(\beta)}=\ket{\psi_{n}}$ can also be written in a recursive fashion as
\begin{align}
\ket{\psi_{k}}=\frac{e^{-\Delta \beta H}\ket{\psi_{k-1}}}{\Vert e^{-\Delta \beta H}\ket{\psi_{k-1}} \Vert }.
\end{align}
By the triangle inequality and Lemma 1 we then find that the error $\epsilon_{k}$ after $k$ steps is upper bounded by  
\begin{align}
\epsilon_{k}&=\Vert  U_{\phi_{k-1}}\ket{\phi_{k-1}}-U_{\psi_{k-1}}\ket{\psi_{k-1}}+U_{\psi_{k-1}}\ket{\psi_{k-1}}- \ket{\psi_{k}}\Vert  \\
\label{eq:splittting}
&\leq \Vert U_{\phi_{k-1}}\ket{\phi_{k-1}}-U_{\psi_{k-1}}\ket{\psi_{k-1}} \Vert +6\Delta\beta^{2}\Vert H\Vert_{\infty}^{2}, 
\end{align}
omitting here the explicit dependence of $U_{\phi_{k-1}}$ and $U_{\psi_{k-1}}$ on $\Delta \beta$.  
We proceed by further upper bounding the first term of the right-hand side. Consider 
\begin{align}\label{eq:SimilarUnitariesBound}
&\Vert U_{\phi_{k-1}}\ket{\phi_{k-1}}-U_{\psi_{k-1}}\ket{\psi_{k-1}} \Vert \\
&=\epsilon_{k-1}+\Delta \beta  \Vert [H,\ket{\phi_{k-1}}\bra{\phi_{k-1}}]\ket{\phi_{k-1}}- [H,\ket{\psi_{k-1}}\bra{\psi_{k-1}}]\ket{\psi_{k-1}} \Vert + R_{\phi_{k-1}}+R_{\psi_{k-1}}\\
&=\epsilon_{k-1}+\Delta \beta \Vert H(\ket{\phi_{k-1}}-\ket{\psi_{k-1}}) -\bra{\phi_{k-1}}H\ket{\phi_{k-1}}\ket{\phi_{k-1}}+\bra{\psi_{k-1}}H\ket{\psi_{k-1}}\ket{\psi_{k-1}} \Vert+ R_{\phi_{k-1}}+R_{\psi_{k-1}}\\
&\leq  \epsilon_{k-1}+\Delta \beta \Vert H\Vert_{\infty}\epsilon_{k-1}+\Delta \beta\Vert \bra{\phi_{k-1}}H\ket{\phi_{k-1}}\ket{\phi_{k-1}}-\bra{\psi_{k-1}}H\ket{\psi_{k-1}}\ket{\psi_{k-1}}\Vert +  R_{\phi_{k-1}}+R_{\psi_{k-1}}
\end{align}
where the remainder terms $R_{\phi_{k-1}}$ and $R_{\psi_{k-1}}$ for the second order of the Taylor expansion of $U_{\phi_{k-1}}$ and $U_{\psi_{k-1}}$ in $\Delta\beta$ can be upper bounded using again the Lagrange form of the remainder to find $R_{\phi_{k-1}}\leq 2\Delta \beta^{2}\Vert H\Vert_\infty^{2}$
and $R_{\psi_{k-1}}\leq 2\Delta \beta^{2}\Vert H\Vert_\infty^{2}$. We thus obtain 
\begin{align}
\Vert U_{\phi_{k-1}}\ket{\phi_{k-1}}-U_{\psi_{k-1}}\ket{\psi_{k-1}} \Vert& \leq (1+\Delta\beta \Vert H\Vert_{\infty})\epsilon_{k-1}+4\Delta \beta^{2}\Vert H\Vert_{\infty}^{2}\\
&+\Delta \beta\Vert \bra{\phi_{k-1}}H\ket{\phi_{k-1}}\ket{\phi_{k-1}}-\bra{\psi_{k-1}}H\ket{\psi_{k-1}}\ket{\psi_{k-1}}\Vert.
\end{align}
Defining $\ket{\Delta_{k-1}} = \ket{\psi_{k-1}} - \ket{\phi_{k-1}}$, we get 
\begin{equation}
\begin{split}
    \bra{\psi_{k-1}}H\ket{\psi_{k-1}}\ket{\psi_{k-1}} =& \bra{\phi_{k-1}}H\ket{\phi_{k-1}}\ket{\phi_{k-1}} + \bra{\Delta_{k-1}}H\ket{\phi_{k-1}}\ket{\phi_{k-1}} 
    \\
    &+ \bra{\psi_{k-1}}H\ket{\Delta_{k-1}}\ket{\phi_{k-1}} + \bra{\psi_{k-1}}H\ket{\psi_{k-1}}\ket{\Delta_{k-1}},
\end{split}
\end{equation}
and therefore
\begin{equation}
    \begin{split}
        \Vert \bra{\phi_{k-1}}H\ket{\phi_{k-1}}\ket{\phi_{k-1}}-\bra{\psi_{k-1}}H\ket{\psi_{k-1}}\ket{\psi_{k-1}}\Vert \leq &
        \left\| \bra{\Delta_{k-1}}H\ket{\phi_{k-1}}\ket{\phi_{k-1}} \right\|+ \left\|\bra{\psi_{k-1}}H\ket{\Delta_{k-1}}\ket{\phi_{k-1}}\right\|
    \\
    & + \left\|\bra{\psi_{k-1}}H\ket{\psi_{k-1}}\ket{\Delta_{k-1}}\right\| \leq 3 \Vert H\Vert_{\infty}\epsilon_{k-1}.
    \end{split}
\end{equation}
We then arrive at 
\begin{align}
\Vert U_{\phi_{k-1}}\ket{\phi_{k-1}}-U_{\psi_{k-1}}\ket{\psi_{k-1}} \Vert \leq (1+4\Delta \beta \Vert H\Vert_{\infty} )\epsilon_{k-1} +4\Delta\beta^{2} \Vert H\Vert_{\infty}^{2}.
\end{align}
With \eqref{eq:splittting} we hence obtain the recursive relation 
\begin{align}
\label{eq:recursion}
\epsilon_{k}\leq \epsilon_{k-1}A+B
\end{align}
where $A=1+4\Delta \beta \Vert H\Vert_{\infty}$ and $B=10\Delta \beta^{2}\Vert H\Vert_{\infty}^{2}$. We simplify the final expressions by further upper bounding $B$ to find an upper bound for the error $\epsilon_{n}$ after $n$ steps given by the geometric series 
\begin{align}
\epsilon_{n}\leq B\sum_{k=0}^{n-1}A^{k}= B \frac{A^{n}-1}{A-1}&\leq  \frac{5}{2}\Delta \beta \Vert H\Vert_{\infty}\left[\left(1+4\frac{\beta}{n}\Vert H\Vert_{\infty}\right)^{n}-1\right]\\
&\leq \frac{5}{2}\Delta \beta \Vert H\Vert_{\infty} \left[e^{4\beta\Vert H\Vert_{\infty}}-1\right],
\end{align}
so long as $\epsilon_{0}=0$,  which completes the proof.  
\qed

\newpage
\subsection{Proof of Theorem 2}\label{app:Theorem2}
In order to prove Theorem 2 of the main paper, we first develop the following Lemma 2. 
\begin{mdframed}
\textbf{Lemma 2}: \textit{The error $\eta_{n}^{(\gamma)}=\Vert \ket{\chi^{(\gamma)}_{n}} - \ket{\phi_{n}}]\Vert$ averaged over a uniform distribution of $\gamma$ after $n$ steps of SRGD with steps size $\Delta \beta=\frac{\beta}{n}$ is upper bounded by}
\begin{equation}
\tilde{b}_{n}= \sqrt{2\Delta\beta\Vert H\Vert_{\infty}}D\left(e^{8\beta \Vert H\Vert_{\infty}}-1\right)^{1/2}
\end{equation}
\textit{and for any $\tilde{\delta}>0$ we have that}
\begin{equation}
    \mathrm{Pr}\left[ \eta_{n}^{(\gamma)} > \tilde{b}_{n} + \tilde{\delta} \right] \leq \frac{2\Delta\beta\Vert H\Vert_{\infty}D^{2}}{\tilde{\delta}^{2}}\left(e^{8\beta\Vert H\Vert_{\infty} }-1\right)
\end{equation}
\end{mdframed}

\textit{Proof}: We define $\ket{\Delta_{k}^{(\gamma)}} = \ket{\chi^{(\gamma)}_{k}} - \ket{\phi_{k}}$ so that the random variable $\eta_{k}^{(\gamma)}=\Vert \ket{\Delta_{k}^{(\gamma)}}\Vert$ is given as the Euclidean norm of the (unnormalized ) ket vector $\ket{\Delta_{k}^{(\gamma)}}$. The error $\eta_{k}^{(\gamma)}$ describes the norm difference between the state $\ket{\chi_{k}^{(\gamma)}}$ prepared by $k$ steps of stochastic Riemannian gradient descent (SRGD) defined by the update rule 
\begin{align}
 \ket{\chi_{k}^{(\gamma)}(s)} = e^{s g_{k}^{(\gamma)} }\ket{\chi^{(\gamma)}_{k-1}}, 
\end{align}
where $g_{k}^{(\gamma)}= D \mathrm{Tr}\left([H, \ket{\chi^{(\gamma)}_{k-1}}\bra{\chi^{(\gamma)}_{k-1}}]iP_{k}\right)iP_{k}$ is the stochastic Riemannian gradient, and the state $\ket{\phi_{k}}$ obtained by $k$ steps of Riemannian gradient descent (RGD) defined by 
\begin{align}
\ket{\phi_{k}(s)} = e^{sG_{k}}\ket{\phi_{k-1}},  
\end{align}
where $G_{k}=[H, \ket{\phi_{k-1}}\bra{\phi_{k-1}}]$ is the Riemannian gradient. The update occurs for a step size $s=\Delta\beta$ and $D$ is the number of orthonormal Pauli basis elements $P_{k}$, satisfying $\text{Tr}(P_{k}P_{k^{\prime}})=\delta_{k,k^{\prime}}$ and $P_{k}^{2}=\frac{\mathds{1}}{d}$, that we sample from.

In general $D=d^{2}-1 = 2^{2N}-1$, but we may pick $D$ to be smaller as long as $G_{k}$ can be decomposed with fewer basis elements. The following proof holds in either case. 

In order to prove Lemma 2, we make use of Jensen's inequality to find 
\begin{eqnarray}
   &\mathbb E_{\gamma}[\eta_{n}^{(\gamma)}]=\mathbb{E}_{\gamma} \| \ket{\Delta^{(\gamma)}_{n}} \| \leq \sqrt{\mathbb{E}_{\gamma} \| \ket{\Delta^{(\gamma)}_{n}} \|^{2}}
    \\
    &\mathrm{Var}(\eta^{(\gamma)}_{n}) = \mathbb{E}_{\gamma}[\eta^{(\gamma) 2}_{n}] - (\mathbb{E}_{\gamma} [\eta^{(\gamma)}_{n}])^{2} \leq \mathbb{E}_{\gamma} \| \ket{\Delta^{(\gamma)}_{n}}\|^{2},
\end{eqnarray}
and thus both the variance and average can be bounded by bounding $\mathbb{E}_{\gamma} \| \ket{\Delta^{(\gamma)}_{n}} \|^{2}$.

The proof proceeds similarly to how Theorem 1 was addressed, in particular we derive something analogous to \eqref{eq:recursion}, but construct the recursion relation in the square $\eta_{k}^{(\gamma) 2}$ of the Euclidean norm of $\ket{\Delta_{k}^{(\gamma)}}$, instead of a recursion relation in the norm. For $\eta_{k}^{(\gamma) 2}$ averaged over the paths $\gamma$ we can perform a Taylor series expansion, up to first order in $s$, to obtain
\begin{equation}
\mathbb{E}_{\gamma}[\eta_{k}^{(\gamma) 2}]  = \mathbb{E}_{\gamma} \braket{\Delta_{k}^{(\gamma)}}{\Delta_{k}^{(\gamma)}} \leq \mathbb{E}_{\gamma}[\eta_{k-1}^{(\gamma) 2}] + \Delta\beta K + \frac{(\Delta\beta)^{2}}{2} R 
\end{equation}
where $K$ is an upper bound of $\frac{d}{ds} \left.\mathbb{E}_{\gamma} \braket{\Delta_{k}^{(\gamma)}}{\Delta_{k}^{(\gamma)}}\right|_{s=0}$ and $R$ is an upper bound for the remainder term $\frac{d^{2}}{ds^{2}} \left.\mathbb{E}_{\gamma} \braket{\Delta_{k}^{(\gamma)}}{\Delta_{k}^{(\gamma)}}\right|_{s=\xi}, \ \xi\in [0,\Delta\beta]$ while the zeroth order term gives $\mathbb{E}_{\gamma}[\eta_{k-1}^{(\gamma) 2}]$ as one would expect.

To determine $K$ we compute
\begin{equation}
\begin{split}
    \left|\frac{d}{ds} \mathbb{E}_{\gamma} \braket{\Delta_{k}^{(\gamma)}}{\Delta_{k}^{(\gamma)}}\right|_{s=0} &= \left|\mathbb{E}_{\gamma}\bra{\phi_{k-1}} \left(G_{k} -g_{k}^{(\gamma)} \right) \ket{\chi^{(\gamma)}_{k-1}} + \text{c.c.}\right|
    \\
    &= \left|\mathbb{E}_{\gamma}\bra{\phi_{k-1}} [H, \ket{\phi_{k-1}}\bra{\phi_{k-1}} - \ket{\chi_{k-1}^{(\gamma)}}\bra{\chi_{k-1}^{(\gamma)}}] \ket{\chi^{(\gamma)}_{k-1}} + \text{c.c.}\right|,
\end{split}
\end{equation}
where ``$\text{c.c.}$'' denotes the complex conjugate and in the 
second line we used that the average $\mathbb E_{P_k}$ in the $k$-the step over the Pauli $P_{k}$ is given by 
\begin{align}
\mathbb E_{P_k}[g_{k}^{(\gamma)}]=[H,\ket{\chi_{k-1}^{(\gamma)}}\bra{\chi_{k-1}^{(\gamma)}}]
\end{align}
the Riemannian gradient $[H,\ket{\chi_{k-1}^{(\gamma)}}\bra{\chi_{k-1}^{(\gamma)}}]$. 
For any anti-hermitian operator $X$ we have $\bra{\phi_{k-1}} X \ket{\chi_{k-1}^{(\gamma)}} + \text{c.c.} = \bra{\phi_{k-1}} X \ket{\phi_{k-1}} + \bra{\phi_{k-1}} X \ket{\Delta_{k-1}^{(\gamma)}} + \text{c.c.} = \bra{\phi_{k-1}} X \ket{\Delta_{k-1}^{(\gamma)}} + \text{c.c.}$, since $\bra{\phi_{k-1}} X \ket{\phi_{k-1}}$ is an imaginary quantity when $X = -X^{\dagger}$.
Therefore, we obtain

\begin{equation}
\begin{split}
    \left|\frac{d}{ds} \mathbb{E}_{\gamma} \braket{\Delta_{k}^{(\gamma)}}{\Delta_{k}^{(\gamma)}}\right|_{s=0} &= \left|\mathbb{E}_{\gamma}\bra{\phi_{k-1}} [H, \ket{\phi_{k-1}}\bra{\phi_{k-1}} - \ket{\chi_{k-1}^{(\gamma)}}\bra{\chi_{k-1}^{(\gamma)}}] \ket{\Delta^{(\gamma)}_{k-1}} + \text{c.c.}\right|
    \\
    &= \left|\mathbb{E}_{\gamma}\bra{\phi_{k-1}} [H, \ket{\Delta^{(\gamma)}_{k-1}}\bra{\phi_{k-1}} + \ket{\chi_{k-1}^{(\gamma)}}\bra{\Delta_{k-1}^{(\gamma)}}] \ket{\Delta^{(\gamma)}_{k-1}} + \text{c.c.}\right| \leq 8 \|H\|_{\infty} \mathbb{E}_{\gamma}[{{}\eta^{(\gamma) 2}_{k-1}}] = K,
\end{split}
\end{equation}
where in the final line, we expressed the difference $\ket{\chi_{k-1}^{(\gamma)}}\bra{\chi_{k-1}^{(\gamma)}}-\ket{\phi_{k-1}}\bra{\phi_{k-1}} =\ket{\Delta^{(\gamma)}_{k-1}}\bra{\phi_{k-1}} + \ket{\chi_{k-1}^{(\gamma)}}\bra{\Delta_{k-1}^{(\gamma)}} $ in terms of $\ket{\Delta_{k-1}^{(\gamma)}}$. We use the Lagrange form of the remainder to determine $R$. Since $\Vert g_{k}^{(\gamma)} \Vert_{\infty}\leq 2D\Vert H\Vert_{\infty}$ and $\Vert G_{k}\Vert_{\infty} \leq 2\Vert H\Vert_{\infty}$ we have 
\begin{align}
\left|\frac{d^{2}}{ds^{2}}\mathbb{E}_{\gamma} \braket{\Delta_{k}^{(\gamma)}}{\Delta_{k}^{(\gamma)}}\right|&=\left |\mathbb{E}_{\gamma} \bra{\phi_{k-1}}e^{-sG_{k}}(G_{k}^{2}+2G_{k}g_{k}^{(\gamma)}+g_{k}^{(\gamma) 2})e^{sg_{k}^{(\gamma)}}\ket{\chi_{k-1}^{(\gamma)}}+\text{c.c} \right| \\
&\leq 32 D^{2} \Vert H\Vert_{\infty}^{2}.  
\end{align}
We thus obtain the recursion relation
\begin{align}
\mathbb{E}_{\gamma}[\eta_{k}^{(\gamma) 2}]\leq \mathbb{E}_{\gamma}[\eta_{k-1}^{(\gamma) 2}]A+B
\end{align}
where $A=1+8\Delta\beta \Vert H\Vert_{\infty}$ and $B=16D^{2}\Delta\beta^{2}\Vert H\Vert_{\infty}^{2}$, which yields after  
$n$ steps the upper bound
\begin{align}
\mathbb{E}_{\gamma}[\eta_{n}^{(\gamma) 2}] \leq 2\Delta\beta \Vert H\Vert_{\infty} D^{2}\left(e^{8\beta \Vert H\Vert_{\infty}}-1\right).
\end{align}
By Jensen's inequality we have established the first part of Lemma 2. The second part can be established using the Chebyshev inequality that states that for a random variable $X$, the probability of being more than $\tilde{\delta}$ away from the mean is upper bounded by 
\begin{equation}
    \mathrm{Pr}\left[ |X - \mathbb{E}[X]| > \tilde{\delta} \right] \leq \frac{\mathrm{Var}(X)}{\tilde{\delta}^{2}} \quad \Rightarrow \quad \mathrm{Pr} \left[ X - \mathbb{E}[X] > \tilde{\delta}\right] \leq \frac{\mathrm{Var}(X)}{\tilde{\delta}^{2}} .
\end{equation}
Considering the random variable $\eta_{n}^{(\gamma)}$ we we thus have 
\begin{equation}
    \mathrm{Pr} \left[ \eta_{n}^{(\gamma)} > \mathbb{E}_{\gamma} \eta_{n}^{(\gamma)} + \tilde{\delta}\right] \leq \frac{\mathrm{Var}(\eta_{n}^{(\gamma)})}{\tilde{\delta}^{2}} \quad \Rightarrow \quad \mathrm{Pr} \left[ \eta_{n}^{(\gamma)} > \tilde{b}_{n} + \tilde{\delta}\right] \leq \frac{\mathrm{Var}(\eta_{n}^{(\gamma)})}{\tilde{\delta}^{2}} \leq \frac{2\Delta\beta\Vert H\Vert_{\infty}D^{2}}{\tilde{\delta}^{2}}\left(e^{8\beta\Vert H\Vert_{\infty} }-1\right) 
\end{equation}
which completes the proof. \qed 

We go on to use Lemma 2 to establish an upper bound the for he average fidelity error $\mathbb E_{\gamma}[\varepsilon_{n}^{(\gamma)}]$ where
$\varepsilon_{n}^{(\gamma)}=1-| \bra{\psi(\beta)} \chi^{(\gamma)}_{n}\rangle|^{2}$
is the fidelity error between a random state created through SRGD and the ITE state.

\begin{mdframed}
\textbf{Theorem 2:} \textit{The average fidelity error  $\bar{\varepsilon}_{n}=\mathbb E_{\gamma}[\varepsilon_{n}^{(\gamma)}]$ after $n$ steps of SRGD with step size $\Delta\beta=\frac{\beta}{n}$ is upper bounded by }
\begin{align}
    b_{n}=\frac{9}{2}\sqrt{2\Delta\beta\Vert H\Vert_{\infty}}D \left(e^{8\beta\Vert H\Vert_{\infty}}-1\right)^{\frac{1}{2}},
\end{align}
\textit{for sufficiently large $n$. For any $\delta>0$, the probability that a random state will give rise to a fidelity error greater than $b_{n}+\delta$ is upper bounded by }
\begin{align}
    \text{Pr}(\varepsilon_{n}^{(\gamma)} > b_{n} + \delta)\leq \frac{8\Delta\beta\Vert H\Vert_{\infty}D^{2}}{\delta^{2}}\left(e^{8\beta\Vert H\Vert_{\infty} }-1\right) 
\end{align}
\end{mdframed}
\textit{Proof}: We first note that 

\begin{equation}
    \epsilon_{n} = \| \ket{\psi(\beta)} - \ket{\phi_{n}}\| = \sqrt{2 - 2\mathrm{Re}(\braket{\psi(\beta)}{\phi_{n}})} \leq \sqrt{2(1- |\braket{\psi(\beta)}{\phi_{n}}|)} \quad \Rightarrow \quad  |\braket{\psi(\beta)}{\phi_{n}}| \geq 1-\frac{\epsilon_{n}^{2}}{2},
\end{equation}
which tells us that
\begin{equation}
    1-|\braket{\psi(\beta)}{\phi_{n}}|^{2} \leq 1-\left(1-\frac{\epsilon_{n}^{2}}{2}\right)^{2} = \epsilon_{n}^{2} - \frac{\epsilon_{n}^{4}}{4} \leq \epsilon_{n}^{2}\leq \sqrt{2}\epsilon_{n},
\end{equation}
where in the last line we made use of the fact that $\epsilon_{n}=\Vert \ket{\psi(\beta)}-\ket{\phi_{n}} \Vert \leq \sqrt{2}$. 
Therefore, 
\begin{equation}\label{eq:Bound_Fidelity}
\begin{split}
\varepsilon_{n}^{(\gamma)} &= 1-| \bra{\psi(\beta)} \chi^{(\gamma)}_{n}\rangle|^{2} = 1-| \bra{\psi(\beta)} \phi_{n}\rangle|^{2} + \braket{\psi(\beta)}{\chi^{(\gamma)}_{n}} \braket{\Delta^{(\gamma)}_{n}}{\psi(\beta)} + \braket{\psi(\beta)}{\Delta^{(\gamma)}_{n}} \braket{\chi^{(\gamma)}_{n}}{\psi(\beta)} - \left|\braket{\psi(\beta)}{\Delta^{(\gamma)}_{n}}\right|^{2}
\\
& \leq \sqrt{2}\epsilon_{n} + 2\|\ket{\Delta_{n}^{(\gamma)} }\| =\sqrt{2}\epsilon_{n} + 2\eta_{n}^{(\gamma)} ,
\end{split}
\end{equation}
where $\epsilon_{n}$ is upper bounded by Theorem 1, and $\eta_{n}^{(\gamma)}$ is characterized by Lemma 2. In the final inequality of \eqref{eq:Bound_Fidelity}, we used that 
\begin{align}
|\braket{\psi(\beta)}{\chi^{(\gamma)}_{n}} \braket{\Delta^{(\gamma)}_{n}}{\psi(\beta)}| \leq \Vert \Delta_{n}^{(\gamma)}\Vert,&~~~~ |\braket{\psi(\beta)}{\Delta^{(\gamma)}_{n}} \braket{\chi^{(\gamma)}_{n}}{\psi(\beta)}|\leq \Vert \Delta_{n}^{(\gamma)}\Vert. 
\end{align}
By Lemma 2 we thus have that the average fidelity error is upper bounded by 
\begin{align}
\bar{\varepsilon}_{n}&\leq  \frac{5}{2} \sqrt{2} \Delta \beta \Vert H\Vert_{\infty}\left(e^{4\beta \Vert H\Vert_{\infty}}-1\right)+ 2\sqrt{2\Delta\beta\Vert H\Vert_{\infty}}D\left(e^{8\beta \Vert H\Vert_{\infty}}-1\right)^{1/2} \\
&\leq \frac{9}{2}\sqrt{2\Delta\beta\Vert H\Vert_{\infty}}D \left(e^{8\beta\Vert H\Vert_{\infty}}-1\right)^{\frac{1}{2}}=b_{n}
\end{align}
where in the last line we assumed that $\Delta\beta\Vert H\Vert_{\infty}\leq 1$, and used the fact that $\sqrt{y^2 - 1} \geq \sqrt{y^2-2y + 1} = (y-1)$ for all $y\geq 1$. This establishes the first part of Theorem 2 for all $n\geq \beta \|H\|_{\infty}$.

We can prove that the distribution we found for $\eta_{n}^{(\gamma)}$ also naturally applies to $\varepsilon_{n}^{(\gamma)}$.
Our goal is to compute the probability that $\varepsilon_{n}^{(\gamma)} > b_{n} + \delta$, so we wish to convert this inequality to the known probability that $\eta_{n}^{(\gamma)} > \tilde{b}_{n} + \tilde{\delta}$.
Noting that $\tilde{b}_{n} = \frac{1}{2}\left(b_{n} - \sqrt{2}\epsilon_{n}\right)$, we see that 
\begin{equation}
    \mathrm{Pr}\left(\varepsilon_{n}^{(\gamma)} > b_{n} + \delta\right) = \mathrm{Pr}\left(\frac{1}{2}\left(\varepsilon_{n}^{(\gamma)} - \sqrt{2}\epsilon_{n}\right) > \tilde{b}_{n} + \frac{1}{2}\delta\right) \leq \mathrm{Pr}\left(\eta_{n}^{(\gamma)} > \tilde{b}_{n} + \tilde{\delta}\right)
\end{equation}
when setting $\tilde{\delta} = \frac{1}{2}\delta$.
The final inequality comes from the fact that $\eta_{n}^{(\gamma)} \geq \frac{1}{2}\left(\varepsilon_{n}^{(\gamma)} - \sqrt{2}\epsilon_{n}\right)$ and 
\begin{equation}
    \mathrm{Pr}\left(\eta_{n}^{(\gamma)} > \tilde{b}_{n} + \tilde{\delta}\right) = \mathrm{Pr}\left(\frac{1}{2}\left(\varepsilon_{n}^{(\gamma)} - \sqrt{2}\epsilon_{n}\right) > \tilde{b}_{n} + \tilde{\delta}\right) + \mathrm{Pr}\left(\eta_{n}^{(\gamma)} > \tilde{b}_{n} + \tilde{\delta} \geq \frac{1}{2}\left(\varepsilon_{n}^{(\gamma)} - \sqrt{2}\epsilon_{n}\right)\right).
\end{equation}
Making use of Lemma 2 we we thus have
\begin{equation}
    \mathrm{Pr}\left(\varepsilon_{n}^{(\gamma)} > b_{n} + \delta\right) \leq \frac{2\Delta\beta\Vert H\Vert_{\infty}D^{2}}{\tilde{\delta}^{2}}\left(e^{8\beta\Vert H\Vert_{\infty} }-1\right) = \frac{8\Delta\beta\Vert H\Vert_{\infty}D^{2}}{\delta^{2}}\left(e^{8\beta\Vert H\Vert_{\infty} }-1\right),
\end{equation}
which completes the proof \qed.

\end{document}